\documentclass[sigconf,nonacm]{acmart}
\usepackage{tabularx}
\usepackage{booktabs}
\usepackage{multirow}
\usepackage{colortbl}
\colorlet{diffrow}{yellow!30}
\colorlet{samerow}{black!7}
\colorlet{absentrow}{red!8}
\usepackage{graphicx}
\usepackage{subcaption}
\graphicspath{ {figures/} }
\DeclareGraphicsExtensions{.pdf,.png,.jpg}

\usepackage{listings}
\usepackage{enumitem}
\usepackage{eurosym}
\lstdefinelanguage{yaml}{
  keywords={true,false,null,y,n},
  keywordstyle=\color{blue}\bfseries,
  basicstyle=\ttfamily\scriptsize,
  sensitive=false,
  comment=[l]{\#},
  morecomment=[s]{/*}{*/},
  commentstyle=\color{gray}\ttfamily,
  stringstyle=\color{teal}\ttfamily,
  moredelim=[l][\color{orange}]{\&},
  moredelim=[l][\color{magenta}]{*},
  morestring=[b]',
  morestring=[b]",
}

\lstdefinelanguage{json}{
  basicstyle=\ttfamily\scriptsize,
  morestring=[b]",
  stringstyle=\color{teal}\ttfamily,
  literate=
    *{0}{{{\color{orange}0}}}{1}
     {1}{{{\color{orange}1}}}{1}
     {2}{{{\color{orange}2}}}{1}
     {3}{{{\color{orange}3}}}{1}
     {4}{{{\color{orange}4}}}{1}
     {5}{{{\color{orange}5}}}{1}
     {6}{{{\color{orange}6}}}{1}
     {7}{{{\color{orange}7}}}{1}
     {8}{{{\color{orange}8}}}{1}
     {9}{{{\color{orange}9}}}{1}
     {:}{{{\color{black}{:}}}}{1}
     {,}{{{\color{black}{,}}}}{1}
     {\{}{{{\color{black}{\{}}}}{1}
     {\}}{{{\color{black}{\}}}}}{1}
     {[}{{{\color{black}{[}}}}{1}
     {]}{{{\color{black}{]}}}}{1},
}

\usepackage{tikz}

\usetikzlibrary{arrows.meta, positioning, fit, calc, backgrounds, patterns}

\usepackage[most]{tcolorbox}

\AtBeginDocument{%
  }

\setcopyright{none}
\copyrightyear{2026}
\acmYear{2026}

\begin{document}

\title{MAS-Lab:
 A Specification‑Driven Validation Framework for Reliable Multi‑Agent Systems}

\author{Jordan Aug\'e}
\email{augjorda@cisco.com}
\authornotemark[1]
\affiliation{%
  \institution{Cisco Systems}
  \city{Paris}
  \country{France}
}

\author{Giovanna Carofiglio}
\email{gcarofig@cisco.com}
\authornotemark[1]
\affiliation{%
  \institution{Cisco Systems}
  \city{Paris}
  \country{France}
}
\author{Giulio Grassi}
\email{gigrassi@cisco.com}
\authornotemark[1]
\affiliation{%
  \institution{Cisco Systems}
  \city{Paris}
  \country{France}
}
\author{Jacques Samain}
\email{jsamain@cisco.com}
\authornotemark[1]
\affiliation{%
  \institution{Cisco Systems}
  \city{Paris}
  \country{France}
}

\renewcommand{\shortauthors}{Aug\'e et al.}

\begin{abstract}

The rapid emergence of LLM-based agentic frameworks has significantly reduced the cost of assembling multi-agent systems (MAS), enabling fast prototyping and exploration of agentic behaviors. However, systems built with current tooling remain ill-suited for reliable, evolvable, and production-grade deployment. In practice, MAS are often developed in an ad-hoc and imperative manner, with agent logic, orchestration, observability, and control tightly interwoven, little to no explicit system-level validation, and development workflows optimized for demonstrations rather than long-lived, governed operation. As a result, behavior observed during experimentation rarely constitutes reliable evidence of behavior in production.

\noindent In this paper, we introduce \emph{MAS-Lab}, a specification-driven framework for principled development and experimental validation of multi-agent systems properties. \textit{MAS-Lab} is designed to transform MAS from collections of scripts into engineered distributed systems by separating semantic intent from operational concerns, making behavior and control explicit, supporting reproducible experimentation, and preserving continuity across lifecycle stages. \textit{MAS-Lab} consists of three layers: a declarative, framework-agnostic agentic specification layer (\textit{Spec}); a stateful MAS Operating System that provides execution and control primitives plugged-in by design (\textit{MAS-OS}); and a set of lab overlays with integrated observability and evaluation tools (\textit{Labs}). Together, these components enable intent-based validation, principled system evolution, and a seamless transition to production-grade MAS.

\end{abstract}

\maketitle

\section{Introduction}

The rapid rise of agentic AI has produced a diverse ecosystem of development frameworks—ranging from workflow- and graph-based systems to agent runtimes and compositional libraries—that significantly reduce the cost and time required to assemble LLM-based multi-agent systems (cfr.~\cite{langchain_docs_2026,langgraph_overview_2026,autogen_2024,microsoft_agent_framework_2026,google_adk_2026,llamaindex_2026,crewai_2024,haystack_agents_2024,dspy_2024}). These frameworks prioritize construction agility, offering modular abstractions for tools, memory, orchestration, and multi-agent interaction, enabling rapid prototyping with minimal infrastructure. Yet as MAS move from demonstrations to real-world deployment, a different set of challenges emerges.
Most systems remain \emph{ad-hoc}, \emph{monolithic}, and \emph{proof-of-concept oriented}, with agent logic, orchestration, observability, and control tightly interwoven in framework-specific code and prompts. Small changes to compositions or coordination patterns often have unpredictable consequences, and experiments are difficult to reproduce across environments or teams. Transitioning from prototypes to production typically requires rebuilding communication primitives, observability pipelines, control mechanisms, and governance infrastructure from scratch. This lack of lifecycle continuity limits reliability, scalability, and evolvability, and means that behavior observed during experimentation rarely provides trustworthy evidence for production.

Recent standardization efforts have begun to structure the ecosystem. The Model Context Protocol (MCP) defines access to tools and data~\cite{anthropic_mcp_2024}; agentic specifications are emerging through OASF and AgentSpec~\cite{oasf,agent_spec_arxiv_2025}; communication layers are being standardized with A2A and Agntcy~\cite{a2a,agntcy}; and tools/skills abstractions are being introduced for LLM-based agents~\cite{anthropic_tools}. The next step is to move beyond interface-level specifications and rethink agentic runtimes and workflows structurally—developing principled foundations for validation, governance, and reliable execution. Our contribution aligns with this shift toward explicit specifications, verifiable behavior, and governance-ready abstractions.

\noindent First, we address today’s systems lack of an explicit specification. Agent roles, workflows, and constraints are buried in prompts and code, making intent opaque and hard to reuse. An evolvable MAS requires a declarative, framework-agnostic description capturing: (i) \emph{semantic intent}—goals, roles, tools, reasoning and interaction patterns, and state; and (ii) \emph{operational constraints}—observability needs, evaluation criteria, and governance controls. Existing formats such as Open Agent Specification and Agent Manifest~\cite{agent_spec_arxiv_2025,agentmanifest,mas_schema} address parts of this need but still describe mostly static structure, leaving behavior and dynamics under-specified.

\noindent Second, we argue for the existence of composable runtime that separates core agentic logic from execution concerns and enables systematic exploration of reasoning or collaboration patterns. Current evaluation frameworks focus on individual agents, tools, or prompts~\cite{agent_eval_survey,helm_eval,toolbench}, but do not assess whether a MAS satisfies global objectives. Observability platforms such as LangSmith, TruLens, and Ragas~\cite{langsmith_observability_2026,trulens_2026,ragas_docs_2026,opentelemetry_docs_2025} remain loosely coupled to agent frameworks and provide no principled link between \emph{intended design} and \emph{observed traces}. As a result, system semantics are entangled with execution environments, and small control-layer changes can silently alter behavior.
This entanglement makes validation inseparable from experimentation. Because intent, execution, and operational constraints cannot be tested independently, reproducibility is low and iteration difficult. MAS research still relies on bespoke harnesses and ad-hoc setups, and even strong benchmarks such as AgentBench and MultiAgentBench~\cite{agentbench_liu_2024,multiagentbench_2025} offer limited reproducibility. Recent systems like MAESTRO~\cite{maestro_2026} improve cross-framework observability but emphasize operational metrics rather than conformance to declared intent, and lack a runtime that evolves with the application lifecycle. Security monitors such as Sentinel Agents~\cite{sentinel_agents_2025} detect violations but do not verify adherence to a declarative MAS specification. Consequently, MAS-level intent, interaction correctness, and emergent behavior remain largely un-validated.

\noindent To address these gaps, we introduce \textit{MAS-Lab}, a specification-driven stack for development, validation and experimentation of multi-agent systems. MAS-Lab treats semantic intent, execution, control, and deployment as first-class, explicitly separated concerns, transforming MAS from collections of scripts into engineered distributed systems.
\textit{MAS-Lab} comprises three layers. The \textbf{Agentic Specification Layer} provides a declarative, framework-agnostic description of MAS intent, constraints, and control requirements, enabling intent-based validation and principled evolution. The \textbf{MAS Operating System (MAS-OS)} offers a modular, framework-independent runtime that interprets specifications and exposes hooks for observability, policy enforcement, coordination, recovery, and adaptation. Finally, the \textbf{Lab Overlays} define reproducible deployment and evaluation environments with integrated observability and measurement tooling, ensuring that validated properties carry through to production.
Together, these layers enable experimentation and validation workflows previously impractical without bespoke infrastructure. MAS-Lab allows design‑space exploration without code changes, carries a single specification across development and production for consistent conformance checks, and ties quality differences to concrete trace events rather than aggregate metrics. It also supports principled assessment of MAS topologies, quantifying both improvements and null results with equal rigor.

\noindent The remainder of the paper is structured as follows: \S\ref{sec:rw} reviews related work, \S\ref{sec:obj} outlines MAS-Lab objectives, and \S\ref{sec:arch} presents the architecture. Based on the sample application in \S\ref{sec:samples} as a running example, \S\ref{sec:spec} and \S\ref{sec:runtime} describe the specification layer and MAS-OS runtime, while \S\ref{sec:mas-lab} and \S\ref{sec:results} cover the lab environment and the evaluation. \S\ref{sec:discussion} discusses implications and limitations, and \S\ref{sec:conclusions} concludes.

\begin{table*}[t]
\centering
\small
\begin{tabular}{l lcccc}
\toprule
\textbf{Category} &
\textbf{System (reference)} &
\textbf{Execution Runtime} &
\textbf{Lifecycle Awareness} &
\textbf{Reproducibility} &
\textbf{Spec-driven Validation} \\
\midrule

\multirow{4}{*}{\textbf{Declarative Spec.}} &
OASF~\cite{oasf}
  & -- & Implicit & -- & Interface-level \\

& AgentSpec~\cite{agent_spec_oracle_2025}
  & Declarative & Implicit & -- & Interface-level \\

& MCP~\cite{mcp_spec_2025}
  & Stateless & -- & -- & Interface-level \\

& AGNTCY~\cite{agntcy}
  & Orchestrated & Operational & -- & Interface-level \\

\midrule
\multirow{6}{*}{\textbf{Agentic Runtime}} &
LangGraph~\cite{langgraph_overview_2026}
  & Stateful, Orchestrated & Implicit & Ad-hoc & -- \\

& AutoGen~\cite{autogen_2024}
  & Orchestrated & -- & Ad-hoc & -- \\

& LlamaIndex~\cite{llamaindex_2026}
  & Stateless & -- & -- & -- \\

& CrewAI~\cite{crewai_2024}
  & Orchestrated & -- & Ad-hoc & -- \\

& Haystack Agents~\cite{haystack_agents_2024}
  & Stateful, Orchestrated & Operational & Ad-hoc & -- \\

& DSPy (agentic use)~\cite{dspy_2024}
  & Stateless & -- & Metric-level & -- \\

\midrule
\multirow{2}{*}{\textbf{Agent OS}} &
Agent OS (Builder Methods)~\cite{agentos_buildermethods}
  & -- & Explicit & Ad-hoc & Tooling-level \\

& agent-os (Rivet)~\cite{agentos_rivet}
  & Isolated & -- & -- & -- \\

\midrule
\multirow{3}{*}{\textbf{Evaluation}} &
AgentBench~\cite{agentbench_liu_2024}
  & Orchestrated & -- & Metric-level & -- \\

& MAESTRO~\cite{maestro_2026}
  & Orchestrated & Operational & Trace-level & -- \\

& MASLab (Ye et al.)~\cite{maslab_ye_2025}
  & Stateful, Orchestrated & -- & Trace-level & -- \\

\midrule
\textbf{This work} &
\textbf{MAS-Lab}
  & \textbf{Declarative, Stateful}
  & \textbf{Explicit}
  & \textbf{End-to-end}
  & \textbf{Design-bound} \\

\bottomrule
\end{tabular}
\caption{Positioning of MAS-Lab w.r.t. related work.}
\label{tab:spec}
\end{table*}

\section{Related Work}
\label{sec:rw}

This section reviews related work across four areas—declarative agent specifications, agentic runtimes, agent operating systems, and MAS evaluation frameworks—and positions \textit{MAS-Lab} with respect to the capabilities and limitations of each.

\subsection{Declarative Agent Specifications}

The \textbf{Open Agentic Schema Framework (OASF)}~\cite{oasf} provides a structured, extensible representation of agent capabilities, metadata, and inter-agent relationships. It offers a semantic substrate for describing agents and MAS structure in a framework-agnostic manner, forming the basis for the declarative layer in \textit{MAS-Lab}. However, OASF focuses on schema interoperability and discovery rather than runtime invariants, coordination semantics, or execution-level validation.
Standardization efforts such as the \textbf{Open Agent Specification (AgentSpec)} introduce declarative descriptions of agents, tools, and workflows with portable execution semantics. While AgentSpec improves component portability, it does not capture MAS-level lifecycle constraints, coordination invariants, or trace-integrated governance. Complementary initiatives such as the \textbf{Model Context Protocol (MCP)} and \textbf{AGNTCY} focus on tool access, capability negotiation, identity, and secure messaging~\cite{anthropic_mcp_2024,a2a,agntcy}. These efforts enhance interoperability but do not define declarative MAS semantics or validation mechanisms.
Earlier \emph{cognitive agent specification languages}, including \textbf{CASL} and \textbf{BDI}-based frameworks, provide rigorous logical models for agent reasoning and coordination. While theoretically strong, they are not aligned with modern LLM-driven agents, tool-centric interaction, or stochastic execution.

\noindent \textit{MAS-Lab} extends OASF with a concrete MAS contract model covering messages, tools, state, and control, together with a design-to-trace validation engine and schema-driven orchestration. It combines AgentSpec-style portability, MCP/AGNTCY interoperability, and the semantic rigor of cognitive agent languages to provide a unified declarative MAS specification enabling explicit system intent and runtime conformance validation.
Beyond specification standards, the MASLab framework by Ye et al.~\cite{maslab_ye_2025} consolidates over twenty LLM-based multi-agent reasoning and coordination methods into a common benchmarking environment. Their work focuses on comparing coordination strategies rather than specifying or validating system-level intent, invariants, or lifecycle constraints. It complements our approach: MASLab standardizes agent methods, while \textit{MAS-Lab} provides a holistic framework for MAS specification, execution, control, and validation.

\subsection{Agentic Runtimes}

A wide range of agentic runtimes manage execution, tool invocation, and coordination for LLM-based agents. Workflow- and graph-based systems such as LangGraph~\cite{langgraph_overview_2026} and AutoGen~\cite{autogen_2024} structure agent interactions through stateful graphs or message-passing conversations. These frameworks offer modular coordination but remain imperative, embedding system intent and invariants directly in orchestration logic.
Production-oriented frameworks such as the Microsoft Agent Framework~\cite{microsoft_agent_framework_2026} and Google’s ADK~\cite{google_adk_2026} provide unified environments for deploying and scaling agents, but do not support declarative specification or system-level validation. Compositional libraries such as LangChain~\cite{langchain_docs_2026} and LlamaIndex~\cite{llamaindex_2026} enable rapid experimentation with tools, memory, and retrieval pipelines, yet leave lifecycle continuity and reproducibility to application-specific implementations.
Across these systems, semantic intent, execution logic, and operational control are typically intertwined, limiting principled validation and safe evolution. 
\noindent In contrast, \textit{MAS-Lab} separates declarative MAS intent from execution and deployment, enabling intent-based validation while remaining interoperable with existing runtimes.

\subsection{Agent Operating Systems}

Several systems adopt the notion of an \emph{agent operating system} to provide persistent structure and control. Builder Methods’ \emph{Agent OS} captures organizational context and project standards to improve agent reliability~\cite{agentos_buildermethods}, but does not define execution semantics or system-level validation. Runtime-oriented systems such as Rivet’s \emph{agent-os}~\cite{agentos_rivet} emphasize sandboxed execution and secure tool invocation through WebAssembly and V8 isolates. These approaches strengthen execution substrates but leave agent intent and coordination semantics implicit, highlighting the need for architectures that bridge declarative specifications with runtime behavior.

\subsection{MAS Experimentation and Evaluation}

Evaluating MAS is challenging due to non-determinism, long-horizon dependencies, and emergent behavior. Benchmarks such as AgentBench~\cite{agentbench_liu_2024} and MultiAgentBench~\cite{multiagentbench_2025} assess task performance and coordination quality but do not support specification or validation of system-level intent.
MAESTRO~\cite{maestro_2026} provides framework-agnostic infrastructure for testing, observability, and reproducible experimentation, standardizing execution and measurement across agent frameworks. Its evaluation model is operational and post hoc, lacking ties to declarative system intent. MASLab~\cite{maslab_ye_2025} standardizes implementations of coordination methods for controlled comparison but does not bind evaluation to design-time invariants.

\noindent In contrast, \textit{MAS-Lab} treats validation and evaluation as architectural consequences of declarative specification. Execution traces and metrics are interpreted relative to explicit MAS contracts, enabling semantic conformance checking, principled analysis of design changes, and preservation of evidence across development, validation, and production.

Table~\ref{tab:spec} positions \textit{MAS-Lab} contribution with respect to related work. In terms of execution runtime, we distinguish: \textit{Stateless}:per-invocation execution with no persistent agent state;\textit{Stateful}: execution maintains durable state across interactions;\textit{Orchestrated}: centralized control manages agent coordination; \textit{Isolated}: agents execute in sandboxed environments; \textit{Declarative}: runtime behavior is driven by external specifications.

\section{\textit{MAS-Lab} Design Objectives}\label{sec:obj}

\textit{MAS‑Lab} spans the entire lifecycle of a multi‑agent system—from development through validation, testing, and production deployment—ensuring that the implemented behavior remains correct, aligned with the specified intent, and compliant with declared control and governance policies.
Its design is led by the following objectives:
\begin{enumerate}

\item \textbf{Separate Semantics from Operations}, by supporting a declarative specification of a MAS that clearly separates logic definition (agent skills, tools,
prompts, coordination) from operational concerns (governance, tracing, cost controls), from infrastructure characteristics.

\item \textbf{Formalize MAS Runtime as a Constraint System}. We
model each agent kernel as a composition of finite Mealy machines (one per concern), bounding the behavior of non-deterministic agents into a finite execution path and enabling verifiable governance guarantees.

\item \textbf{Define standard contracts} that make the boundary between an agent (or MAS) and its environment explicit. Clear boundaries are essential for testing, correctness, responsibility attribution, and preserving identity across deployments. Following a POSIX-like philosophy—specifying interfaces without constraining internal design—\textit{MAS-Lab} exposes the agent kernel through a small set of contract interfaces. Some face the agent logic (e.g.\ the \emph{design pattern} and \emph{context} assembly); the others mediate every interaction with the external world: \emph{model} inference plus the four action contracts \emph{tool}, \emph{state}, \emph{communication}, and \emph{control}.

\item \textbf{Built-in Observability}, embedded into the runtime.
The runtime is instrumented to record a complete, ordered event stream at every boundary crossing---tool calls, LLM requests, delegation events, governance decisions, so enabling further analysis, attribution and comparison.

\item \textbf{Reproducibility Artifacts}. Based on recorded events and trajectories, every claim about system behavior is backed by an auditable artifact. \textit{MAS-Lab} aims at providing a complete
experimental harnesses enabling independent validation of
architectural claims and of applied governance.

\end{enumerate}

\section{\textit{MAS-Lab} Architecture}
\label{sec:arch}

\begin{figure}[t]
  \centering
    \includegraphics[width=\columnwidth]{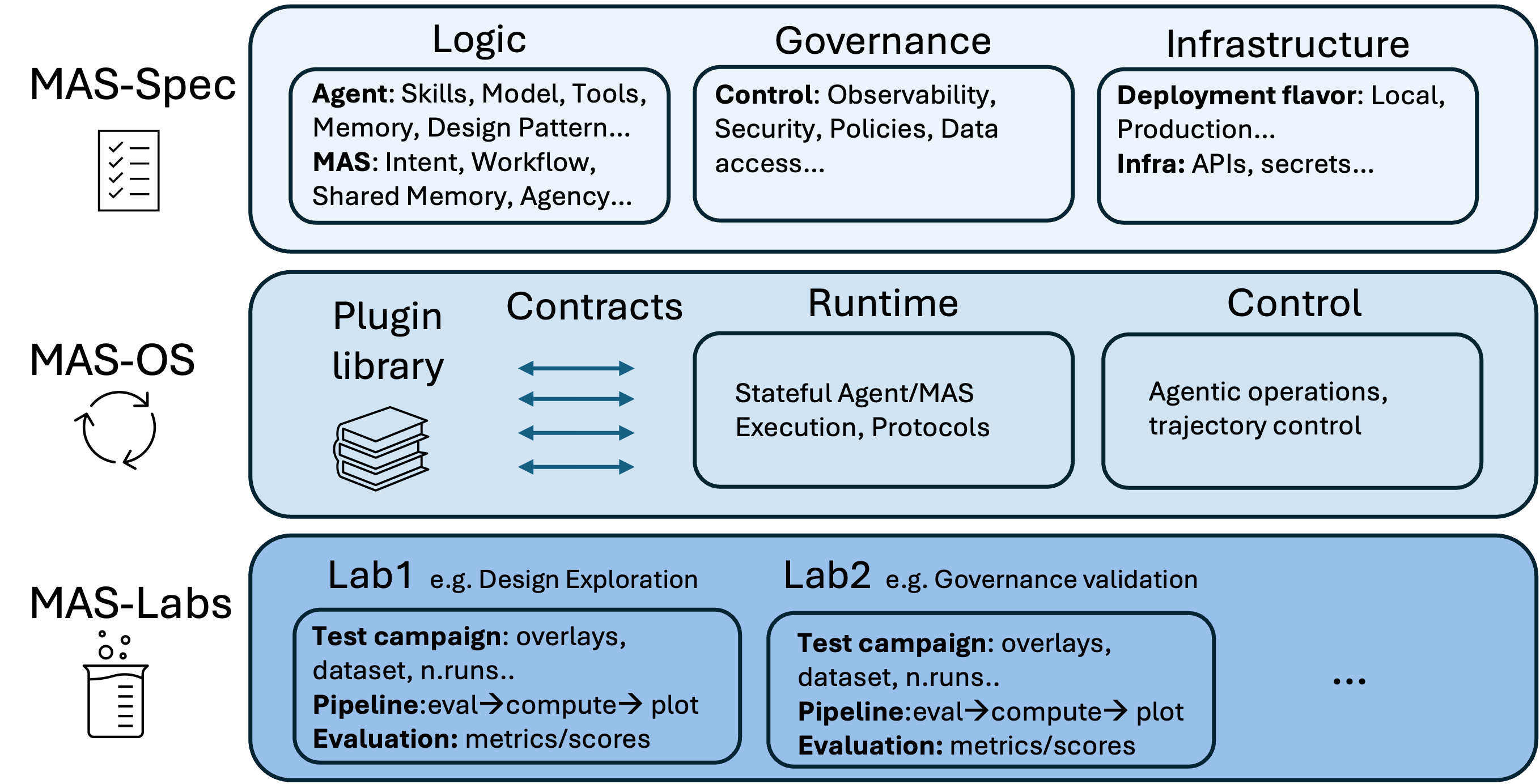}
  \caption{\textit{MAS-Lab} layers.}
  \label{fig:three-layers}
\end{figure}

Based on the design objectives in Sec.~\ref{sec:obj}, \textit{MAS-Lab} is organized into three layers (Figure~\ref{fig:three-layers}): 
\textit{MAS-Spec}, a declarative specification of MAS logic, governance, and infrastructure; 
\textit{MAS-OS}, a contract-based runtime that executes agents through declared bindings; 
and \textit{MAS-Labs}, a set of overlays and pipelines for systematic testing, evaluation, and experimentation.

\subsection{MAS-Lab layers}
\label{sec:layers}

The \emph{agentic specification} (\S\ref{sec:spec}) provides the naming foundation. It declares all system entities—agents, tools, models, and coordination edges—in versioned YAML specifications. Nothing can be executed or evaluated unless it is first declared here.

The \emph{agent runtime} (\S\ref{sec:runtime}) implements these declarations through contracts at every boundary. Tool calls, LLM requests, delegations, and memory accesses are checked against bound plugins, and each event is recorded. Undeclared tools are rejected, undeclared delegations are blocked, and all LLM calls are logged identically across deployments.

The \emph{lab layer} (\S\ref{sec:results}) uses the same specification and runtime to study system behavior under controlled variation. It can swap infrastructure, inject faults, replay traces, or simulate failures—without modifying agent logic or relaxing any contract. Because contract handling is consistent, observed differences are attributable to the experimental variable rather than stack drift.

Across all three layers, the principle is the same: \emph{what is specified can be validated}. Concerns must be declared before the runtime can bind them, and recorded before they can be tested. The specification makes concerns explicit; the runtime materializes them through contracts; the lab exercises them.

The layers share a common vocabulary—specs, contracts, overlays—so that the same artifact describing a system can also be used to test, validate, and compare it.

The remainder of this section details the three layers (\S\ref{sec:spec}--\S\ref{sec:results}).

\subsection{Illustration via a sample MAS}
\label{sec:samples}

We ground the architecture description in a Trip Planner application that exercises the full MAS-Lab stack, from declarative specification to lifecycle control. The planner runs in a fictional city universe backed by deterministic tools (timetables, fare tables, and a routing graph database), removing LLM variability at the tool layer and isolating the effects of agent-design choices. A moderator receives user queries and delegates them to three specialists: a schedule agent for transport timetables and attractions, an itinerary agent for graph-based route planning, and a concierge agent for fare estimation and budget tracking.
Queries span a controlled range of complexity---from simple single-agent lookups (``What time is the next train to Verdania?'') to full-chain coordination requiring all four agents (the moderator plus the three specialists) and, when the Lab~2 governance overlay is applied, budget enforcement
Table~\ref{tab:trip-planner} summarizes the application through its specification, and a runnable tutorial is included in the open-source release accompanying the paper. The Trip Planner serves as the running example for the specification in \S\ref{sec:spec} and the experimental evaluation in \S\ref{sec:results}.

\begin{table*}[ht!]
\centering
\renewcommand{\arraystretch}{1.2}
\begin{tabular}{p{2.8cm} p{14.0cm}}
\toprule
\textbf{Spec Field} & \textbf{Trip Planner MAS (v1.0.0)} \\
\midrule

\textbf{mas\_id} &
\texttt{mas.trip\_planner.v1}. System for planning trips in the Arborian Network. \\

\textbf{purpose / objective} &
Plan trips end-to-end: city selection, schedules, routes, fares, and itinerary assembly. \\

\textbf{agent\_set} &
\texttt{moderator}, \texttt{schedule\_agent}, \texttt{itinerary\_agent}, \texttt{concierge\_agent}.  
Generalist agent available as single-agent baseline. \\

\textbf{orchestration\_flow} &
Moderator is entry point; delegates to schedule, itinerary, and concierge agents.  
Full-chain coordination for complex queries. \\

\textbf{tool\_registry} &
\texttt{lookup\_schedule}, \texttt{get\_attractions}, \texttt{get\_attractions\_description},
\texttt{query\_graph\_database}, \texttt{get\_fares}, \texttt{get\_trip\_fares}, \texttt{get\_attraction\_fare}, \texttt{calc}. \\

\textbf{runtime\_invariants} &
Deterministic tool backends; consistent delegation boundaries; itinerary must match returned schedules, routes, and fares. \\

\textbf{conversation\_policy} &
Moderator uses structured reasoning and explicit delegation; specialists return grounded schedule, route, or fare data. \\

\textbf{governance\_controls} &
Not embedded in the base MAS; applied via Patch overlays. The app ships an HITL-on-tool overlay; budget, content-filter, and circuit-breaker overlays are provided with the Lab~2 lifecycle-control experiment. \\

\textbf{observability\_schema} &
All tool calls, delegations, and LLM events recorded by MAS-OS runtime. \\

\textbf{failure\_injection} &
Labs can inject tool faults, replay traces, or simulate routing/schedule inconsistencies. \\

\textbf{deployment/portab.} &
Runs across flavours: local dev, CI with mock LLMs, production. Same MAS logic across environments. \\
\midrule

\textbf{moderator\_agent}& design pattern: react; moderator-based workflow (entry node); LLM: azure/gpt-4o ($T{=}0.3$);
Tools: none --- delegates to the three specialists via workflow edges (\texttt{delegates\_to});
Intent: classify request complexity, dispatch, assemble itinerary.\\

\textbf{schedule\_agent} & schedules + attractions; LLM: azure/gpt-4o ($T{=}0.2$);
Tools: lookup\_schedule, get\_attractions, get\_attractions\_description;
Intent: return departures, travel times, attraction details.\\

\textbf{itinerary\_agent}& route planning; LLM: azure/gpt-4o ($T{=}0.2$);
Tools: query\_graph\_database, get\_fares;
Intent: compute feasible routes and compare travel times/costs.\\

\textbf{concierge\_agent}& fares + budget; LLM: azure/gpt-4o ($T{=}0.2$);
Tools: get\_trip\_fares, get\_attraction\_fare, query\_graph\_database, calc;
Intent: compute fare breakdowns and track budget.\\

\textbf{generalist\_agent}& single-agent baseline (standalone app, not part of the four-agent MAS); design pattern: react; LLM: azure/gpt-4o ($T{=}0.3$);
Tools: lookup\_schedule, query\_graph\_database, get\_fares, calc;
Skills: trip-orchestration, transport-schedule-lookup, route-planning, fare-and-itinerary-assembly;
Intent: full workflow without delegation. \\
\bottomrule
\end{tabular}
\caption{MAS/agents logic specification for the Trip Planner MAS (v1.0.0).}
\label{tab:trip-planner}
\end{table*}


\section{MAS Specification}
\label{sec:spec}

The first layer of \textit{MAS-Lab} is a declarative, framework-agnostic specification that describes a multi-agent system before any code executes. Every agent, tool, model, skill, and coordination edge is explicitly declared, and every runtime identifier resolved into a named entity. This specification enables attribution, validation, and comparison across all layers of \textit{MAS-Lab}.

Compared to existing specification efforts (\S\ref{sec:rw})—including OASF, AgentSpec, MCP, and AGNTCY—our contribution lies in making governance first-class and in cleanly separating MAS logic from the governance and infrastructure applied at deployment. This separation allows the logic to be validated independently of the operational constraints layered on top of it.

Prior work typically bundles structural metadata, execution portability, and protocol definitions into a single description of a MAS and its implementation. Logic, execution details, and governance are intertwined, and governance itself is largely absent: budgets, sandbox policies, and authorization rules are hardcoded or configured externally, making them invisible to validation, comparison, or reproducibility.

In \textit{MAS-Lab}, governance is declarative. A \texttt{Patch} overlay can specify tool-call budgets, capability toggles, delegation constraints, or context-freshness requirements, versioned alongside the MAS they constrain. These constraints are enforced precisely at the boundary where they apply—not as post-hoc checks, but as contract violations that prevent the boundary crossing.

Observability is also part of the declared governance. A \texttt{Flavour} specification specifies the deployment environment, including where events are recorded. Because every boundary crossing—tool calls, LLM requests, delegations, and governance decisions—is intercepted by construction, any backend becomes natively observable without additional instrumentation. \textit{MAS-Lab} ships with a comprehensive native JSON backend, and an OpenTelemetry exporter interface. Others could be added by plugging on the same contract interfaces.

\subsection{Specification Fields}
\label{sec:speccontent}

The specification is a composition of MAS/agents logic,  governance overlays, and infrastructure choices, merged at deployment time. Table~\ref{tab:trip-planner} illustrates these fields concretely for the Trip Planner MAS.

\noindent\paragraph{Agents/MAS specification.}
A \texttt{MAS} specification declares the system identity (versioned name, tags), a free-text intent, and the agency topology: agents, their roles, and the delegation edges forming the workflow graph. Each agent is defined in a separate \texttt{Agent} specification that declares its models, tools, design pattern, and system prompt. Agents may reference typed skills with I/O contracts and acceptance criteria, making capabilities independently testable. Tools are declared by reference, grounding tool access in specifications rather than ad-hoc code.
\noindent Listing~\ref{lst:manifest} shows the Trip Planner MAS specification and a governance overlay side by side: the base specification declares identity and topology, while the overlay adds constraints without modifying the logic.
\noindent\paragraph{Flavours and infrastructure.}
A \texttt{Flavour} specification selects the operational configuration for a deployment: LLM endpoints, communication protocol, tool access policies, caching, and telemetry. Flavours and infrastructure specifications describe the concrete resources and policies used by a deployment. The same MAS can run in local development, CI with mock LLMs, or production against real endpoints without modification.
\noindent\paragraph{Overlays.}
A \texttt{Patch} overlay adds governance or experimental conditions without altering the base MAS. Patches can enable or disable memory, restrict tool budgets, enforce delegation policies, or set context-freshness requirements. Scenario overlays define test conditions—budget overruns, privilege violations, constraint conflicts—as declarative YAML rather than ad-hoc scripts. Composition order is: base MAS, then scenario overlay, then flavour, with CLI flags overriding all.

\subsection{From Specification to Contracts}
\label{sec:spec-to-contracts}

Each declared requirement binds to a runtime contract (\S\ref{sec:runtime}). The specification is the source of truth; the runtime resolves those bindings on the per-agent kernel. Tool declarations bind to the \texttt{ToolContract}; governance constraints (budgets, content filters, approvals) to the \texttt{GovernanceContract}; delegation edges and topology to the \texttt{OrchestrationContract}; context contributions and freshness to the \texttt{ContextContract}; and telemetry to the observability backends declared in the \texttt{Flavour}.

Validation occurs before execution (structural checks), during execution (contract enforcement), and after execution (trace analysis). A system that passes all three stages is operating within its declared intent—or produces an auditable record of every deviation.

\begin{lstlisting}[language=yaml,basicstyle=\ttfamily\scriptsize,
  caption={Trip planner logic (left); governance Patch (right).},label=lst:manifest]
# base MAS specification
apiVersion: mas/v1
kind: MAS
metadata:
  name: trip-planner
spec:
  intent: "Plan multi-city trips in the Arborian Network"
  agents:
    - id: moderator
      role: broker
      delegates_to: [schedule_agent, itinerary_agent, concierge_agent]
    - id: itinerary_agent
      tools: [query_graph_database, get_fares]
    - id: concierge_agent
      tools: [get_trip_fares, get_attraction_fare,
              query_graph_database, calc]
---
# governance Patch overlay
apiVersion: mas/v1
kind: Patch
metadata:
  name: budget-governance
spec:
  target: { kind: MAS, name: trip-planner }
  patches:
    - path: agents.concierge_agent.budget
      value: { max_cost_usd: 0.50, max_tokens: 4000 }
    - path: agents.itinerary_agent.require_human_approval
      value: true   # query_graph_database egress
                     # requires explicit operator approval
\end{lstlisting}

\section{MAS-OS}
\label{sec:runtime}

An agent operates through an iterative cycle of reasoning and action: it constructs inputs for a model, processes outputs, and may invoke tools, access or update state, or delegate tasks. These interactions are inherently asynchronous and non-deterministic, due to latency, rate limits, token streaming, and delayed callbacks that evolve independently of internal control flow. In parallel, internal processes—policy enforcement, orchestration, and tracing—also progress. Execution may be paused, constrained, or blocked while awaiting external results. The runtime must therefore maintain authoritative control over state transitions, ensuring consistency with declared constraints despite asynchronous events.

The requirements of §\ref{sec:obj} make this separation essential. External interactions must be preceded by explicit policy evaluation and produce records tied to that evaluation, rather than post hoc logs. Control actions (pause, abort, approval) must prevent late completions without corrupting state. Backends—model providers, tool transport, telemetry—must remain configurable without modifying agent logic. Experiments must reuse the same specification and produce evidence linked to it. \textit{MAS OS} provides a runtime satisfying these constraints from development to production.

Hook-based frameworks typically conflate control and I/O within a single execution stack, making it difficult to interpose decisions between intent and effect: pauses or denials may fail, and late responses may still affect state. \textit{MAS OS} addresses this by separating control from execution.

The \textbf{kernel} is the deterministic control core, advancing state through a totally ordered sequence of named events, without network I/O or provider dependencies. The \textbf{execution engine} performs impure operations—HTTP, MCP, filesystem, messaging—only at designated execution steps after authorization. The kernel delegates work and later ingests outcomes as events. While execution workers may run concurrently, the kernel enforces a total event order, enabling consistent replay and audit.

Ensuring policy-before-execution requires an explicit representation of control state (e.g., awaiting model output, validating a tool request, paused for an operator). Implicit coordination mechanisms (flags, callbacks) do not remain coherent under concurrency. Finite-state control instead makes states and transitions explicit, ensuring that system evolution occurs only through declared events.

Among finite-state models, automata classify inputs but do not bind decisions to execution steps. Moore machines associate outputs with states, whereas Mealy machines associate outputs with transitions. Mealy semantics better capture authorization, as decisions (allow, deny, defer) are attributed to specific events. More expressive controllers increase complexity and reduce composability. The kernel is therefore implemented as a composition of finite Mealy machines, one per concern, preserving finiteness while separating tool interaction, model calls, lifecycle, policy, and tracing.

These machines do not run as independent threads. Instead, the kernel maintains a single ordered event queue. At each step, it emits one event from a fixed vocabulary observed by all machines. Each machine updates only if a transition is defined; otherwise it remains unchanged. The global control state is the tuple of local states. A single component—the declarative orchestration binding—proposes the next interaction, while others react to the resulting events.

Consider a guarded tool call such as the itinerary agent's \texttt{query\_graph\_database} lookup under a human-in-the-loop approval overlay. The orchestration schedules the interaction and the tool machine enters validation. Upon authorization, multiple machines (tool, governance, approval, recorder) update simultaneously: approval may defer execution, the tool enters a waiting state, and the recorder logs the decision. Execution is deferred until a subsequent event resumes the interaction. Asynchronous completions re-enter as events. The engine performs I/O only at execution steps, while the kernel advances deterministically one event at a time.

A complete formal treatment is beyond scope. The following subsections describe how specifications bind to the per-agent Mealy kernel (§\ref{sec:runtime-declared}), how policy governs external work (§\ref{sec:runtime-governed}), how kernels coordinate in multi-agent systems (§\ref{sec:runtime-mas}), and how traces are tied to the specification (§\ref{sec:runtime-traces}).

\subsection{Specification bindings on the kernel}
\label{sec:runtime-declared}

\textit{MAS-Lab} targets multi-agent systems, but execution is always realized through a \textbf{per-agent kernel}: a single ordered Mealy control history associated with one agent identity defined in the versioned specification (§\ref{sec:spec}). 
A multi-agent application is therefore a set of kernels composed through explicit topology, rather than a single loop encoding multiple roles in prompts.
The specification declares agents, tools, state stores, communication edges, and policies. At runtime, these declarations are resolved into contract plugins bound to the kernel’s control charts. We focus here on a single kernel; coupling across kernels is described in §\ref{sec:runtime-mas}.

\begin{table}[t]
\centering
\small
\begin{tabular}{@{}p{0.22\linewidth}p{0.34\linewidth}p{0.36\linewidth}@{}}
\toprule
\textbf{Scope} & \textbf{What runs} & \textbf{Primary artifact} \\
\midrule
Single agent &
  One kernel, one timeline, one agent id &
  Per-agent trace \\
MAS (specification) &
  Named agents, edges, policies &
  Versioned manifest \\
MAS (runtime) &
  Several kernels, routed communication &
  Coordinated traces \\
\bottomrule
\end{tabular}
\caption{Declaration vs.\ runtime scope (composition in Section~\ref{sec:runtime-mas}).}
\label{tab:execution-scope}
\end{table}

Every outward action falls into one of five classes: \textbf{model} inference, \textbf{tool} invocation, \textbf{state} access, \textbf{communication}, and \textbf{execution} control (pause, cancel, resume, spawn). All classes follow the same interaction pattern: open, govern, record, execute (if allowed), record, close. Context assembly occurs prior to these steps and is treated as internal preparation rather than a separate outward action.
Kernel transitions are of two types: \textbf{hooks} and \textbf{execution transitions}. Hooks invoke contract logic (e.g., policy checks, payload preparation) and update internal state without external I/O. Execution transitions occur only after authorization and are the sole points where external actions (API calls, messaging, I/O) are performed via the execution engine. Recording spans both: every transition is observed and logged, so traces capture intermediate state changes as well as final outcomes.
The kernel itself is fixed for a given runtime release; variability is introduced exclusively through the specification. 
\textbf{Agent logic} (persona, tools, orchestration, delegation) is versioned with the specification to preserve interpretability of traces. 
\textbf{Reusable control} (budgets, approvals, guardrails, recorders) is defined once and applied across agents. 
\textbf{Infrastructure bindings} map abstract interfaces (models, tools, transport, telemetry) to concrete backends. A given interaction produces the same trace whether backed by a local stub or a remote service; only the binding differs.

Runtime-facing services follow the same principle. Recording may target JSON logs or OpenTelemetry without changing interaction semantics. Execution control may rely on local signals or distributed coordination services. In all cases, behavior is defined by the protocol, not by the backend.
Policies and governance are expressed declaratively and enforced through \textbf{contracts}. Each concern (tool use, communication, recording) is encapsulated behind a contract interface, allowing local or protocol-based implementations (e.g., MCP, A2A, OTel) without modifying agent logic. This decoupling allows logic, policy, and infrastructure to evolve independently.
Two external actors interact with the kernel through the same guarded interface: the \textbf{world} (models, APIs, peers) and the \textbf{user/operator}. All effects are mediated through declared interactions; application code cannot bypass this protocol.
Because specification and execution are decoupled, the same declared system supports both production execution and controlled experimentation. Varying a single specification dimension (e.g., policy, backend, orchestration) while keeping the interaction model fixed enables comparisons whose differences are attributable to the declared change rather than implementation artifacts.

\subsection{Policy before external work}
\label{sec:runtime-governed}

\begin{figure*}[t]
\centering
\small

\noindent\textbf{Governed \texttt{query\_graph\_database} when the operator chooses \emph{reject}}\\[0.35em]

\begin{tikzpicture}[
  >=Stealth,
  bx/.style={draw, rounded corners=2pt, minimum width=1.35cm,
             minimum height=0.85cm, font=\scriptsize, align=center, inner sep=2pt},
  skipped/.style={draw=gray!40, dashed, fill=gray!6, text=gray!45!black},
  active/.style={draw=gray!55, fill=white},
  cause/.style={draw=orange!80!black, line width=1.2pt, fill=orange!12},
]
  \node[bx, active, fill=blue!8] (s1) at (0,0.85) {%
    Call\\ opens\\[-1pt]{\tiny trace start}};
  \node[bx, cause] (s2) at (2.15,0.85) {%
    Governance\\ before tool\\[-1pt]{\tiny policy check}};
  \node[bx, skipped] (s3) at (4.3,0.85) {%
    Trace\\ update};
  \node[bx, skipped] (s4) at (6.45,0.85) {%
    Tool\\ runs\\[-1pt]{\tiny external I/O}};
  \node[bx, skipped] (s5) at (8.6,0.85) {%
    Trace\\ update};
  \node[bx, skipped] (s6) at (10.75,0.85) {%
    Governance\\ on result};
  \node[bx, active, fill=blue!8] (s7) at (12.9,0.85) {%
    Call\\ closes\\[-1pt]{\tiny trace end}};

  \draw[->, thick, gray!45, dashed] (s1) -- (s2);
  \draw[->, thick, gray!45, dashed] (s2) -- (s3);
  \draw[->, thick, gray!45, dashed] (s3) -- (s4);
  \draw[->, thick, gray!45, dashed] (s4) -- (s5);
  \draw[->, thick, gray!45, dashed] (s5) -- (s6);
  \draw[->, thick, gray!45, dashed] (s6) -- (s7);

  \node[font=\tiny, text=gray!50, anchor=south, align=center] at (6.45,1.55)
    {Happy path (grey dashed): used when governance allows the tool to run};

  \node[bx, draw=green!50!black, fill=green!8, minimum width=6.0cm,
        minimum height=1.35cm] (pause) at (4.3,-1.15) {%
    \textbf{Paused --- operator review}\\[2pt]
    Run is \texttt{PAUSED}. Operator chooses \textbf{reject} (or \textbf{approve}).\\
    Reject is written into context for the next LLM call.\\
    \textbf{Route lookup not executed;} no tool result added.};

  \draw[->, very thick, orange!80!black] (s2.south) -- (pause.north)
    node[midway, right=0.12cm, font=\scriptsize, align=left]
    {Policy requires\\operator input};

  \draw[->, very thick, green!50!black] (pause.east) -| (s7.south)
    node[near start, above, font=\scriptsize, align=center]
    {Resume, reject call,\\close call};

  \node[font=\tiny, align=center, text=red!70!black] at (6.45,-2.45)
    {On decline the tool never runs---blocked by governance, not by a tool failure};
\end{tikzpicture}

\vspace{0.55em}

\setlength{\tabcolsep}{4pt}
\renewcommand{\arraystretch}{1.18}
\footnotesize
\begin{tabular*}{\textwidth}{@{\extracolsep{\fill}} l p{0.52\textwidth} p{0.34\textwidth} @{}}
\toprule
\textbf{Stage} &
\textbf{What happens} &
\textbf{Trace (observability)} \\
\midrule
Model proposes \texttt{query\_graph\_database} &
Main scheduler enters the tool phase; the LLM call completes. &
Tool intent recorded. \\

Tool call opens &
Tool handler starts validation; no external query yet. &
Tool-call trace opened. \\

Governance before tool &
Governance matches \texttt{query\_graph\_database} and requires operator confirmation. &
Approval required for \texttt{query\_graph\_database}. \\

Run paused &
Execution control moves to \texttt{PAUSED}; scheduler and tool handler wait. &
Pause recorded. \\

Operator chooses reject &
Operator submits reject (boolean or menu choice vs.\ approve). &
Operator choice recorded. \\

Context updated &
Context assembly stores the decline for the next LLM call. &
Context commit recorded. \\

Run resumed &
Execution control returns to \texttt{RUNNING}. &
Resume recorded. \\

Tool rejected &
Runtime rejects the tool call; the database is never queried. &
Rejection recorded; no tool execution event. \\

Tool call closes &
Call finishes without a route-lookup result. &
Tool-call trace closed. \\

Next model call &
Scheduler prepares context assembly and a new LLM call. &
New LLM cycle trace opened. \\

\bottomrule
\end{tabular*}

\caption{%
  Trip Planner \texttt{query\_graph\_database}, operator chooses \textbf{reject}.
  \emph{Top:} standard call pattern; orange box is governance \emph{before}
  the tool runs; green path records the operator choice in context and closes
  the call without contacting the database backend.
  \emph{Bottom:} same story as a readable timeline with trace entries.%
}
\label{fig:mealy-trace}
\end{figure*}
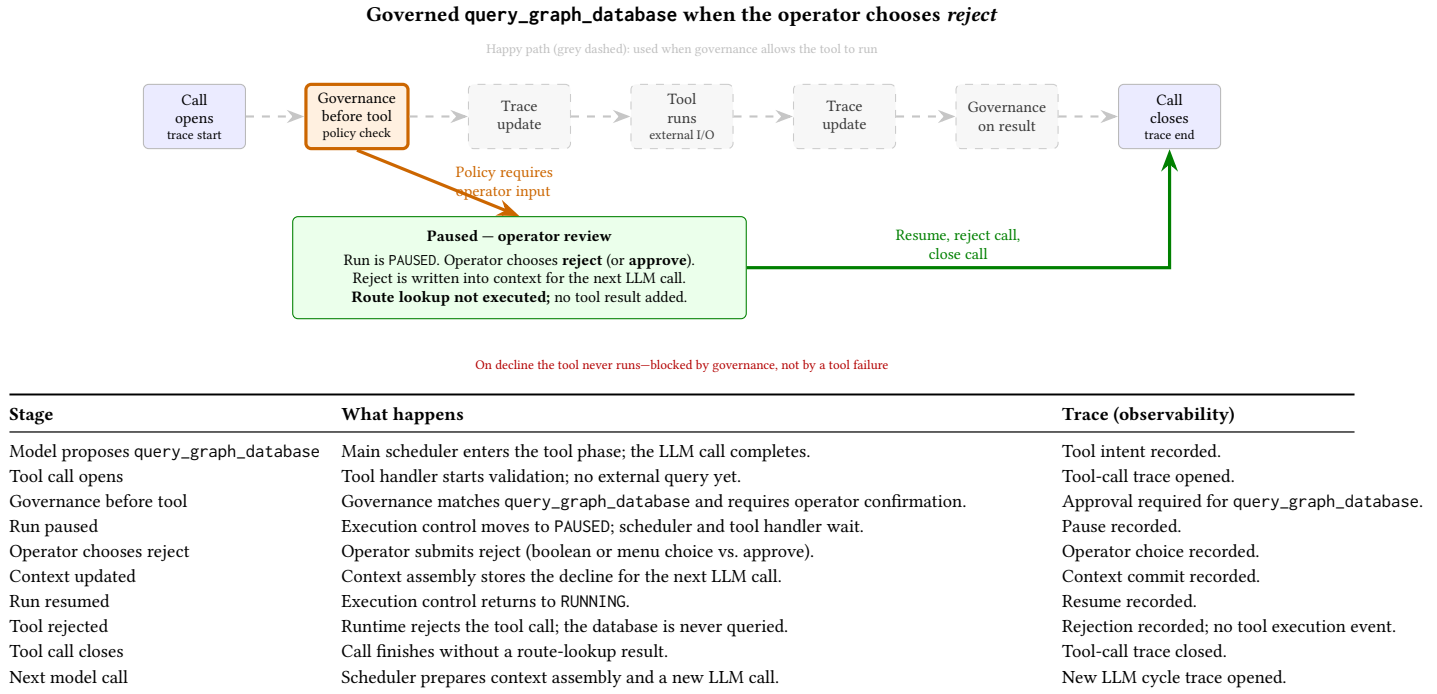

All outward interactions follow a common pattern: open, govern, record, execute (if allowed), record, close. Figure~\ref{fig:mealy-trace} illustrates this sequence for tool calls; the same structure applies to model inference, state access, communication, and execution control.
The governance step ensures that no external action occurs without prior authorization. For example, Listing~\ref{lst:manifest} requires operator approval for the itinerary agent's \texttt{query\_graph\_database} call. When the model proposes this call, the kernel opens the interaction but halts at governance before any external invocation.

Figure~\ref{fig:mealy-trace} illustrates the reject path. Policy matches \texttt{query\_graph\_database} and defers the interaction pending operator input. The system pauses; upon rejection, the decision is recorded and incorporated into context. The interaction is then closed without executing the tool, and the system transitions to the next model step. The absence of the external call is explicit in the trace, rather than inferred from missing logs.
Throughout the interaction, hooks implement policy and recording, while execution transitions are taken only when authorized. As a result, every decision—allow, defer, reject—is explicitly recorded and causally linked to subsequent behavior.


This controlled interaction model applies uniformly across all kernels. Section~\ref{sec:runtime-mas} extends it to multi-agent compositions.

\subsection{Multi-agent coordination}
\label{sec:runtime-mas}

Multi-agent execution is realized as a composition of per-agent kernels rather than a single global controller. Each agent maintains its own ordered event timeline and applies the interaction rules described above. The specification defines the \textbf{topology}: agents, communication edges, allowed interactions, and transports.
In the Trip Planner example, the moderator and specialists (schedule, itinerary, concierge) are independent kernels. Delegation occurs through explicit \textbf{communication interactions}: the sender performs policy checks, records the send, and hands off execution to the transport. The receiving agent processes the message as a new event on its own timeline and may initiate further interactions.

It is important to distinguish \textbf{handoff} from \textbf{agent instantiation}. Handoff is a communication interaction between existing kernels, whereas starting a new agent is an execution action that creates a new kernel instance. Both are governed and recorded but represent distinct events.
This architecture supports composition with external or third-party agents. Interaction is constrained at the protocol boundary, so policy enforcement and traceability remain local, even when peers are opaque.
Although kernels may run across processes or hosts, each kernel advances one event at a time. Distributed execution replaces in-memory communication with transport protocols, but does not alter control semantics. Correlation identifiers link events across kernels, enabling end-to-end trace reconstruction.
The Trip Planner also illustrates why MAS cannot be modeled as a single global state machine. Agents progress independently; cross-agent dependencies emerge through communication rather than shared state. Synchronization properties therefore arise from kernel coupling, not from individual interactions.

\subsection{Runtime traces and correlation}
\label{sec:runtime-traces}

Every kernel transition—hook or execution—produces a trace event recording the interaction, applied policies, participating components, and outcomes. These traces are interpreted in the context of the versioned specification that defines agent identity, capabilities, and constraints.
This enables precise reasoning about system behavior. For example, the absence of a tool execution can be attributed to a policy decision recorded at governance time, rather than inferred from missing logs.
In multi-agent systems, correlation identifiers link events across kernels into coherent end-to-end traces. Analysis therefore shifts from interpreting model outputs to reconstructing causal execution paths: which agent performed which action, under which policy, and in what order.
These traces serve as primary evidence for operation and evaluation. Replay, fault injection, and backend substitution modify execution conditions but preserve trace semantics. As a result, experiments remain reproducible and directly attributable to the declared system specification.

\section{Labs: Testing, Benchmarking, and Evaluation Environment}
\label{sec:mas-lab}

MAS-Lab provides a controlled environment for evaluating multi-agent system (MAS) behaviour under a wide range of conditions. Because the runtime is composed of modular contracts and pluggable components, experiments can instrument every layer that influences execution—agent design, control plane, and infrastructure access—rather than treating the system as an opaque black box. This enables fine‑grained \emph{what‑if} studies, including checkpoint–resume execution, interception of platform access, deterministic replay of LLM outputs, and simulation of tool or service failures. A specification that would normally run on Kubernetes with MCP servers and agent‑to‑agent (A2A) messaging can instead be executed locally on a single‑process communication bus to validate agentic logic in isolation. This separation decouples functional correctness from the variability introduced by network protocols and serving infrastructure, making it practical to scale tests that focus on reasoning, coordination, and reliability rather than deployment mechanics.

Within this environment, MAS-Lab supports three overlapping communities. Application developers use it to explore the design space of an agentic application, validate agent logic independently of deployment infrastructure, and iterate on performance under reproducible conditions. Researchers benefit from a platform that enables systematic evaluation of new mechanisms—design patterns, control policies, governance layers—under comparable execution environments, with the ability to instrument every boundary crossing and run controlled perturbation studies. Enterprise operators rely on MAS-Lab to assess resilience, governance conformance, and operational correctness by injecting failures, replaying traces, and validating behaviour before production rollout. Together, these capabilities make MAS-Lab a unified foundation for principled development, rigorous experimentation, and reliable deployment of multi-agent systems.

A central capability of MAS-Lab is its declarative benchmark framework, which defines, executes, and post‑processes experiments using the same specification vocabulary as the runtime itself. Experiments become versioned, resumable artefacts: partial runs can be restarted from the last completed step, and baseline scenarios can be reused across studies through a content‑addressed run cache keyed on the effective application specification.

\begin{figure}[t]
  \centering
  \includegraphics[width=\linewidth]{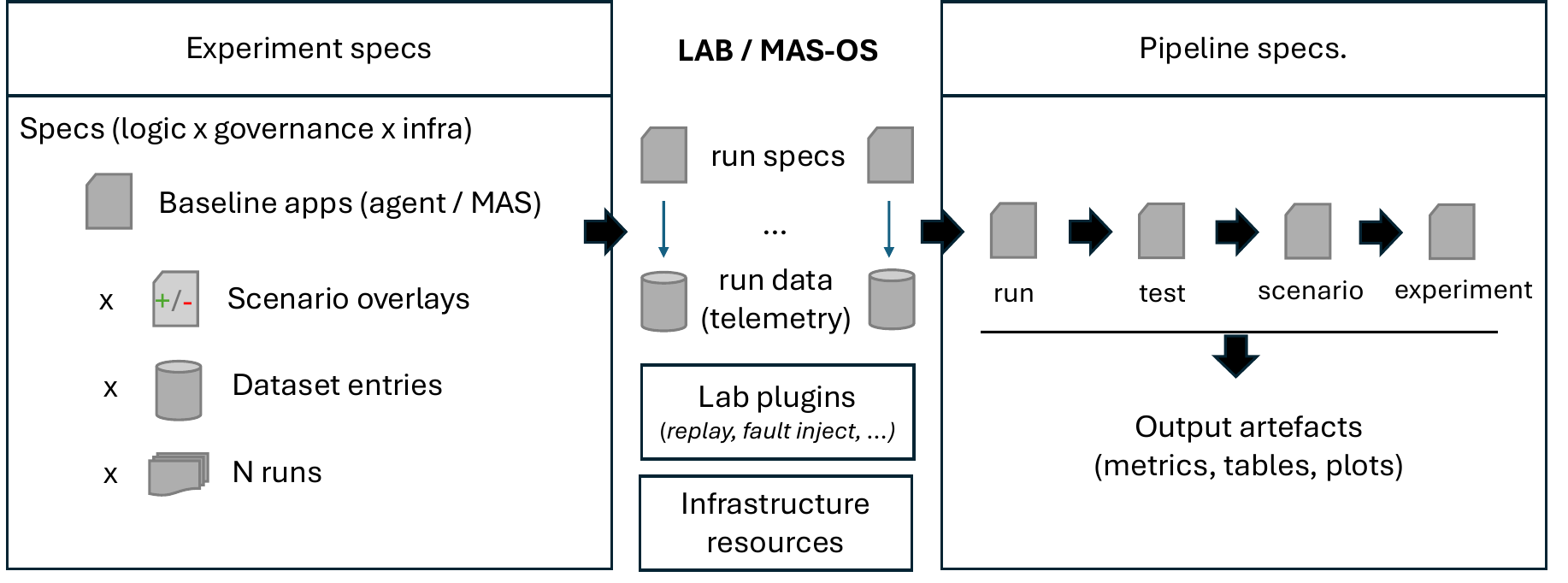}
  \caption{MAS-Lab flow: from experiment specifications through MAS-OS execution and control to
    Labs pipeline steps producing results artefacts.}
  \label{fig:mas-lab-overview}
\end{figure}

MAS-Lab provides a controlled environment for evaluating multi-agent system (MAS) behaviour under a wide range of conditions. Its \textbf{controlled execution and observability} model exposes privileged access to agent internals through a contract-based hook plane, allowing plugins and lab components to attach to lifecycle events—before and after LLM calls, tool invocations, and control operations—to record trajectories, enforce policy, inject faults, or substitute cached responses. Layered \emph{resource emulation} further separates concerns: infrastructure backends (LLMs, tools, memory) and runtime transport (local bus versus network protocols) can be configured independently, enabling experiments that replay cached model outputs while exercising live tool stubs, or vice versa. Together, these mechanisms turn MAS-Lab into an instrumented testbed rather than a passive runner: the same MAS specification can be studied under nominal conditions, under controlled degradation, or under fully deterministic replay for regression analysis.

Building on this foundation, MAS-Lab defines a structured \textbf{experiment model} that mirrors the dimensions typically varied in agentic system benchmarks. Four nested lifecycle levels scope both execution and pre- or post-processing hooks. At the top level, the \emph{application} identifies the agent or MAS under test, referenced by its specification; several application specifications may be included when comparing topologies that share agent definitions. Each application contains one or more \emph{scenarios}, which represent controlled variants—often single-axis ablations—expressed either as distinct MAS specifications or, more commonly, as overlay documents that patch only the attributes under study. Scenarios contain \emph{tests}, which bundle user prompts, multi-turn dialogues, human-in-the-loop replies, and environmental fixtures such as memory seeds or tool mocks. Finally, each test fans out into multiple \emph{runs} to isolate nondeterminism introduced by LLM sampling; each run carries a seed that propagates consistently across caching and emulation subsystems to prevent accidental cross-seed reuse.


Table~\ref{tab:experiment-levels} summarises the scope of each level and typical pipeline responsibilities.

\begin{table}[t]
  \small
  \centering
  \begin{tabular}{@{}llp{4.5cm}@{}}
    \toprule
    Level & Scope & Typical post-processing \\
    \midrule
    Application & Whole experiment & Aggregate metrics, comparative plot \\
    Scenario    & One overlay column & Scenario-level plots, ablation tables \\
    Test        & One dataset item     & Per-item trajectory analysis \\
    Run         & Single run index     & Trace export, span extraction \\
    \bottomrule
  \end{tabular}
    \caption{Experiment lifecycle levels in MAS-Lab.}
  \label{tab:experiment-levels}
\end{table}

The \textbf{execution engine, caching, and pipelines} ensure that experiments run efficiently and reproducibly. The benchmark scheduler executes scenarios until all specified runs complete and their artefacts are available for analysis, and execution is idempotent: interrupted studies resume from the last successful step rather than restarting from scratch. A content-addressed run cache hashes the effective application specification (base specification plus overlays), allowing identical scenarios—such as shared baseline columns—to be reused across experiments without redundant LLM calls. Post-processing is expressed as composable pipelines: ordered steps that consume and produce typed artefacts such as traces, span collections, metric tables, and plots. Pipelines may be referenced from a shared library, loaded from files, or declared inline in the experiment specification. \texttt{pre:} steps at the application level are suited to provisioning external collectors or databases—for example, OpenTelemetry backends—while \texttt{post:} steps transform raw run outputs into analysis-ready forms. MAS-Lab ships with predefined pipeline steps sufficient to reproduce the empirical results of this paper.

\begin{figure}[t]
  \centering
  \includegraphics[width=\linewidth]{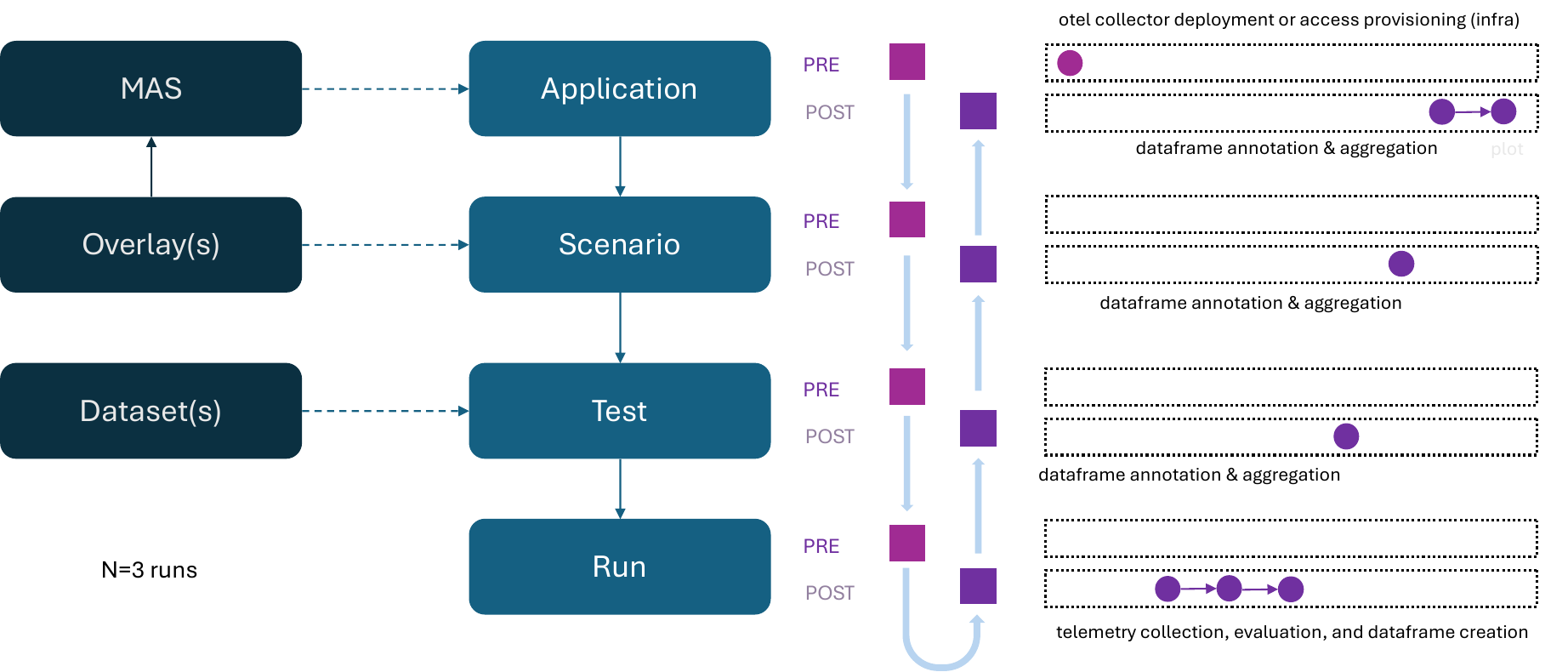}
  \caption{Typical post-execution pipeline flow.
    Typed artefacts flow through reusable steps;
    evaluation and visualisation steps are decoupled from execution.}
  \label{fig:benchmark-pipeline-flow}
\end{figure}

Evaluation is performed through \textbf{MCE}, the Metrics Computation Engine~\cite{telemetryhub2025}, an open-source analysis framework that provides native trajectory metrics and adapters for LLM-as-judge providers. MCE consumes OpenTelemetry spans exported during runs and computes scores such as goal success rate, groundedness, and response completeness. Because evaluation reuses the same infrastructure configuration as the experiment itself, state-of-the-art evaluators can be applied transparently without bespoke glue code, and caching can be leveraged for LLM-as-judge requests.

Finally, MAS-Lab emphasises \textbf{interoperability and outlook}. Beyond immediate reproducibility, it serves as a vendor-neutral substrate for sharing agentic mechanisms and the benchmarks that validate them. The runtime’s contract interfaces define an interoperability layer: new design patterns, control mechanisms, reliability techniques, and evaluation engines can be packaged as library plugins together with declarative experiment specifications, datasets, and published results. Coupled with extensive observability and flexible emulation, the platform supports systematic exploration at both the single-agent and MAS levels. The following section illustrates these capabilities on concrete case studies and details the contract model that enables hook-level instrumentation and cross-study reuse.

\section{Experimental Evaluation}
\label{sec:results}

The objective of this section is to demonstrate---through small, controlled
experiments---which MAS-Lab features are practically usable for exploratory and
benchmarking workflows.  The three labs cover successive stages of an agentic system
lifecycle: \textbf{Lab~1} addresses \emph{design space exploration} via the question about reasoning and
topology selection for a given task execution; \textbf{Lab~2} addresses
\emph{agentic control} illustrating how \textit{MAS-Lab} helps adding and testing operational safeguards without changing the
validated agentic logic code); \textbf{Lab~3} addresses \emph{agentic context exchange} for proper semantic and cognitive behavior by exploring how knowledge
sources contribute to overall outcome.
Each Lab builds upon the previous one and shows incremental steps in the evolution of the agentic system from design to operational deployment.
The numerical results serve as evidence that the framework answers each
question correctly; they are not the primary claim.
All experiment specifications, scenarios, datasets, and analysis pipelines are released as
open-source lab implementations~\cite{outshiftopen2025}.

\subsection{Experimental Setup}
\label{sec:eval-setup}

\paragraph{Application and dataset.}
\textit{MAS-Lab} is released with two reference applications used in this paper: a
\emph{single-agent} QA assistant and a \emph{multi-agent} Trip Planner
(\S\ref{sec:samples}).
Lab~1 / Exp~1.1 (design pattern comparison) uses the single-agent app, evaluated on a 100-item
reasoning-query dataset, while other experiments use the Trip Planner application:
Lab~1 / Exp~1.2 (topology comparison), Lab~2 (lifecycle control), and Lab~3
(context-source extensions).

Datasets differ by experiment, and all are released in \texttt{library-samples}.
Exp~1.1 uses a 100-item single-agent reasoning-query dataset; Exp~1.2 uses a
100-item Trip Planner benchmark. Lab~2 uses a smaller lifecycle dataset of six
nominal queries plus fault-injection items, and Lab~3 uses a 10-item
fact-recall dataset grouped by the number of personal facts the answer
requires. Where a dataset distinguishes query difficulty (e.g.\ simple
single-constraint vs.\ complex multi-hop items), results are reported
separately for the two groups when applicable.

\paragraph{Baseline and run cardinality (Lab~1).}
Lab~1 is defined as overlays applied to fixed baseline specification:
the baseline keeps application logic and tool backends unchanged, while each
variation changes exactly one declared factor.
For Exp~1.1 (single-agent QA), we evaluate 5 design-pattern variations on a
100-item dataset, with 1 run per variation-item pair
($5 \times 100 \times 1 = 500$ runs).
For Exp~1.2 (Trip Planner), we evaluate 5 topology variations on a 100-item
dataset, with 1 run per variation-item pair
($5 \times 100 \times 1 = 500$ runs).


\paragraph{Evaluation framework.}

For quality evaluation we leverage the Metrics Computation Engine
(MCE)~\cite{telemetryhub2025}, an open-source agentic evaluation framework that computes metrics from OTel observability telemetry.
The MCE uses a plugin architecture that integrates third-party evaluation
frameworks (DeepEval~\cite{deepeval}, Opik, RAGAS) alongside native agent metrics.
For answer quality we report the \emph{Answer Relevancy} from DeepEval, and \emph{Goal Success Rate} (GSR); a native
LLM-as-a-Judge score indicating whether the response fulfills the
stated query objective.
The MCE runs as a post-execution pipeline step; it reads from the OTel span
store populated by the \texttt{ObserveSDKPlugin} during each run.
No evaluation code is added to agents.

\paragraph{Infrastructure and reproducibility.}
All runs use \texttt{azure/gpt-4o} model unless otherwise noted.
Infrastructure is held constant across conditions within each experiment (same LLM endpoint, same tool backends).
Trace caching is declared at the infrastructure level: the infra specification
specifies a content-addressed trace store, and the execution engine skips
conditions whose (specification, dataset item, overlay) fingerprint matches a cached
trace.
This makes incremental experimentation efficient: adding a scenario to an
existing experiment re-executes only the new condition.
All result tables in this section are generated automatically by declared
pipeline steps; no numbers are transcribed manually.

\subsection{Lab~1: Design Space Exploration}
\label{sec:lab1}

A developer building a MAS needs to make early choices about reasoning protocol
(chain-of-thought, plan-execute, reflection, etc.) and agents' collaboration topology
(parallel, pipeline, moderator, etc.).  These choices profoundly affect
outcome quality, cost and latency, but are not easy to derive from the problem assigned to agents and thus require to be experimentally assessed.  In traditional MAS frameworks, testing which choice
is better requires code forks, refactoring, or embedded conditional logic ---
and when you measure performance across different implementations, you cannot
isolate whether a latency or quality difference comes from the design choice
itself or from subtle implementation differences, random variation, or data
artifacts.  
The core problem is: \emph{how do I measure the contribution of
one design decision while holding everything else constant?}

Lab~1 demonstrates that this problem becomes tractable with declarative
overlays.  The same agent specification runs under five different reasoning
patterns (Exp~1.1) and five different topologies (Exp~1.2), with only the
overlay differing.  No agent code is modified; the execution contracts,
telemetry schema, and evaluation pipeline remain fixed.  This ensures each run
differs only on the intended design axis.  Execution traces are
content-addressed (fingerprinted by specification, dataset item, and overlay),
so repeated conditions pull from cache instead of re-running.  A deterministic
post-run pipeline (\texttt{eval\_mce} $\rightarrow$ \texttt{collect\_metrics}
$\rightarrow$ \texttt{compute\_ci} $\rightarrow$ plotting) computes confidence
intervals from traces without modifying code, producing auditable figures from
the same data.

We ran Exp~1.1 on a 100-item reasoning dataset (five patterns: CoT, ReAct,
Plan-and-Execute, Reflection, Tree-of-Thoughts) and Exp~1.2 on trip-planner
items (five topologies: parallel, pipeline, moderator-broker, supervised,
verifier).  Both measure goal success rate, answer relevancy, and latency.

\begin{figure}[htbp]
\centering
\includegraphics[width=\columnwidth]{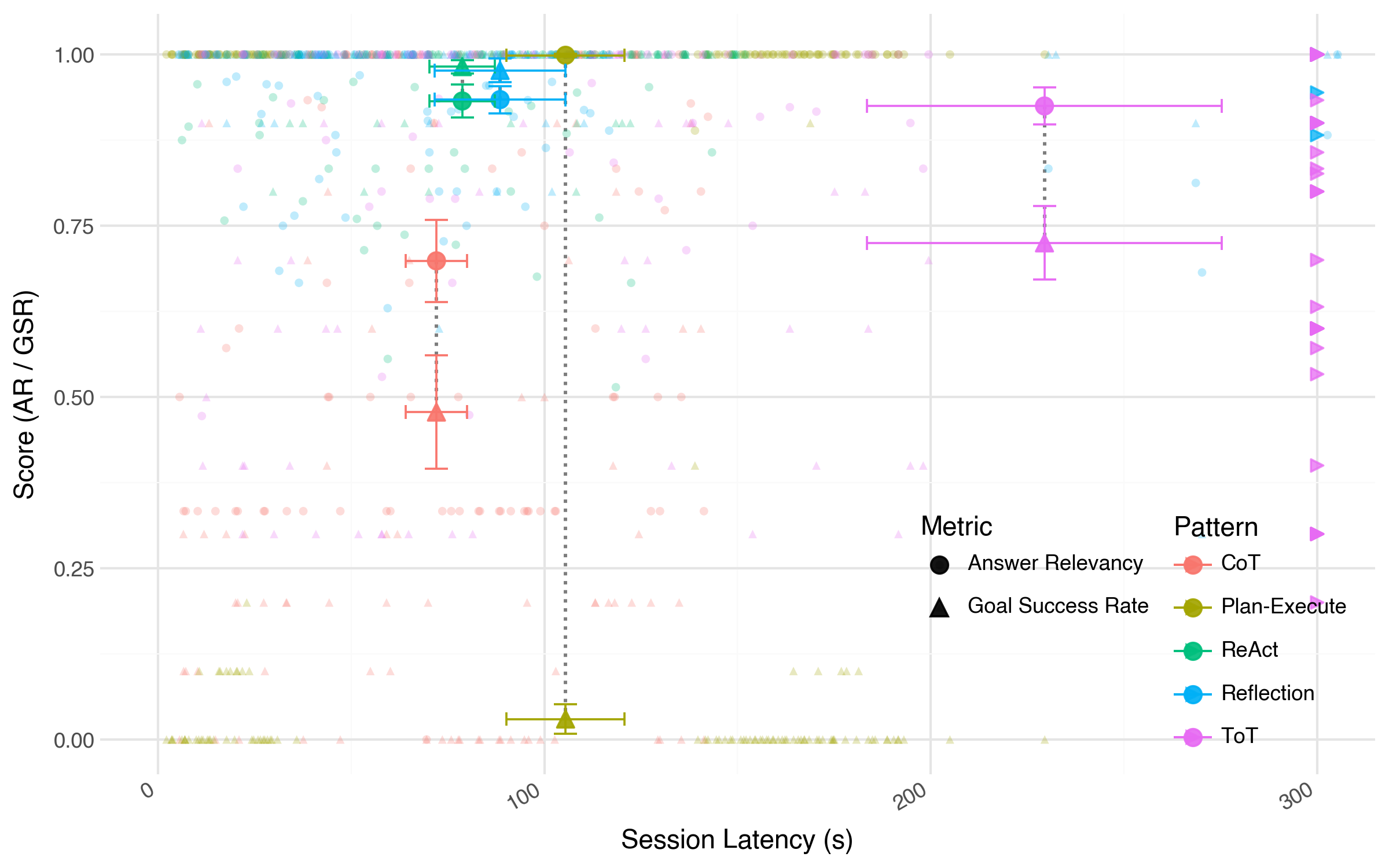}
\caption{Lab~1, Exp~1.1 (reasoning patterns): latency-- quality score trade-off.
  Light points are individual runs; large markers are per-scenario means with
  95\% confidence intervals on both axes.}
\label{fig:lab1-patterns}
\end{figure}

\begin{figure}[htbp]
\centering
\includegraphics[width=\columnwidth]{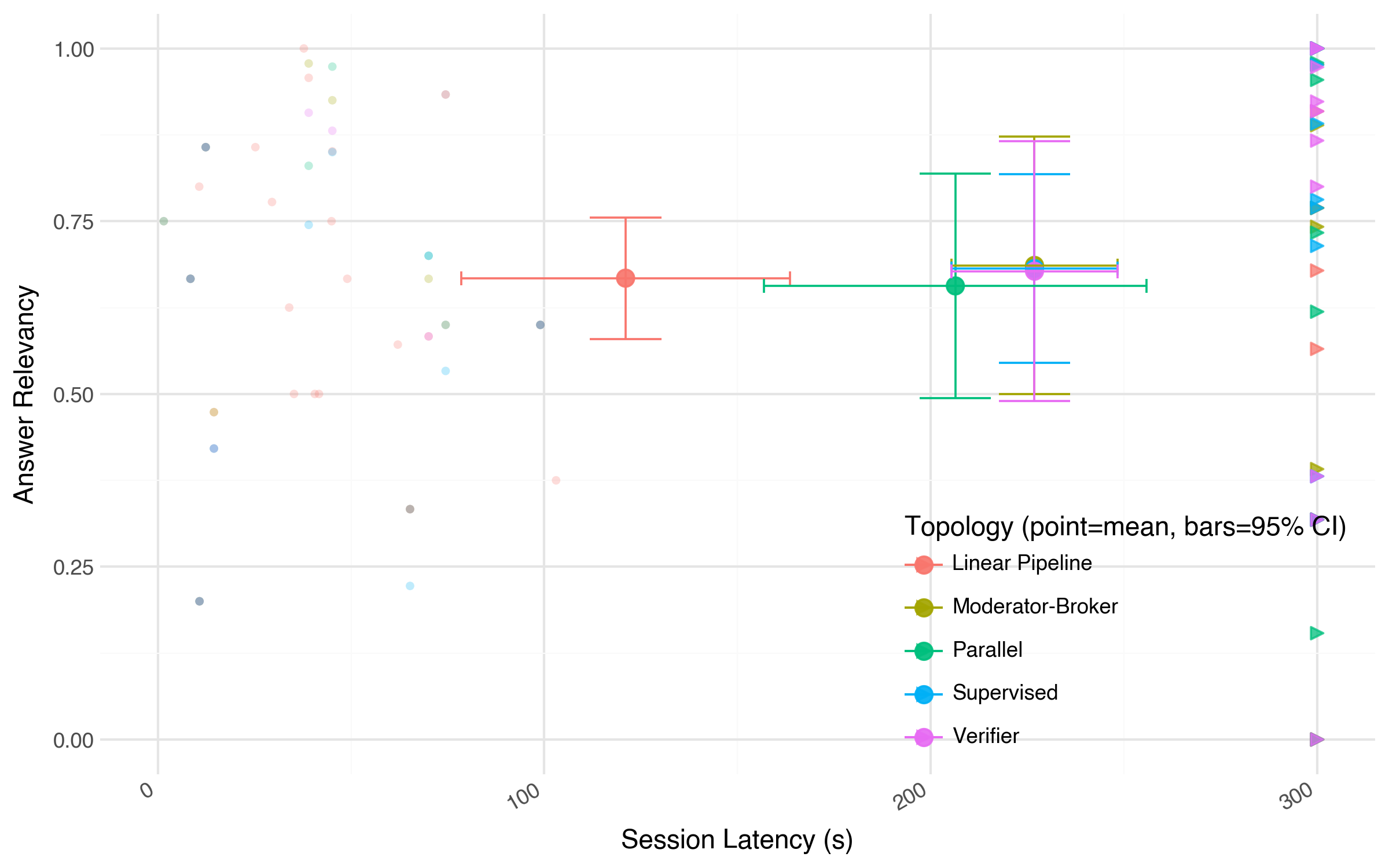}
\caption{Lab~1, Exp~1.2 (workflow topologies): latency-- quality score trade-off for
  Answer Relevancy (\textbullet) and Goal Success Rate (\textbf{$\blacktriangle$}).
  Light points are individual runs; large markers are per-scenario means with
  95\% confidence intervals on both axes.
  Gray dotted lines connect the AR and GSR means for each topology.}
\label{fig:lab1-topology}
\end{figure}

Figures~\ref{fig:lab1-patterns} and \ref{fig:lab1-topology} show each design
choice's isolated effect.  On this dataset, reasoning strategies produce
distinct latency--quality signatures; one pattern may achieve high accuracy at
the cost of latency, another may be fast but less precise.  An important
observation: Plan-and-Execute reaches near-perfect text quality (AR~0.999) but
near-zero task completion (GSR~0.030), a dissociation visible only when you
plot both metrics.  This is not a universal claim; it is specific to our
dataset and task.  Larger datasets or different domains may reveal different
patterns.

What \emph{is} universal is the methodology: a developer can now measure the
contribution of a single design choice by changing one YAML field, without code
refactoring.  The measurement is reproducible because traces cache and the
pipeline is deterministic.  This converts design exploration from an engineering
problem (fork code, manage branches) into a measurement problem (edit specification,
run pipeline, read figures).

Now that design-space exploration is instrumentable, new questions become
answerable: How do reasoning patterns interact across different task
categories?  Do topologies scale consistently as agent count increases?  What
is the joint effect of pattern and topology on a single task?  These questions
require the infrastructure Lab~1 establishes: isolated variation, deterministic
evaluation, and trace-based metrics.  MAS-Lab makes systematic variation
possible; the research questions are now open to developers and researchers who
want to explore their own design spaces.

Lab~2 takes the complementary step: once a design is chosen and validated in
isolation, it must be deployed with operational controls.  The question becomes
whether adding those controls changes what was already validated.


\subsection{Lab~2: Lifecycle Control}
\label{sec:lab2}

When deploying to production, teams add controls on top of an already-validated
agent: observability, budget caps, content filters, circuit breakers.
The requirement is that these additions do not alter what the agent does when
they do not fire, and that when they do fire, their action is safe and
predictable.
For this experiment we implemented four simple governance plugins, each
enforcing one policy type (Table~\ref{tab:lab2controls}).
In a real deployment, each plugin would be subject to its own test suite
before being composed onto the agent; the experiment here treats the plugins as
correct and focuses on whether composing them changes agent behaviour and how
much they cost.
Because each plugin is an independent component with no access to agent state,
the expected property is that the agent's execution trace is identical to the
baseline whenever no policy condition is met --- exactly as if the controls
were not there.
To verify this, all conditions share a response cache: a plugin intercepts
outbound LLM requests at the infra layer and returns the cached answer for any
message it has seen before, regardless of which overlay is active.
Any change in what the agent does is therefore attributable exclusively to a
control firing.

\begin{table*}[htbp]
\caption{Lab~2 governance plugins. Each enforces a declared condition at a
  specific lifecycle point, before the underlying operation runs. The trigger
  is evaluated deterministically against execution state; no probabilistic
  judgement is involved.}
\label{tab:lab2controls}
\small
\begin{tabular}{lllll}
\toprule
\textbf{Plugin} & \textbf{Trigger condition} & \textbf{Lifecycle point} & \textbf{Action} & \textbf{Fault injected} \\
\midrule
Observability   & always                                       & every event           & record span  & --- \\
Budget cap      & cumulative cost $>$ threshold                & before each LLM call  & block + log  & threshold set below actual cost \\
Content filter  & tool argument contains blocked value         & before each tool call & deny + log   & query destination set to ``Shadowmere'' \\
Circuit breaker & $N$ consecutive tool failures                & after each tool call  & halt + log   & tool forced to fail $N$ times \\
\bottomrule
\end{tabular}
\end{table*}

\begin{figure}[htbp]
  \centering
  \includegraphics[width=\columnwidth]{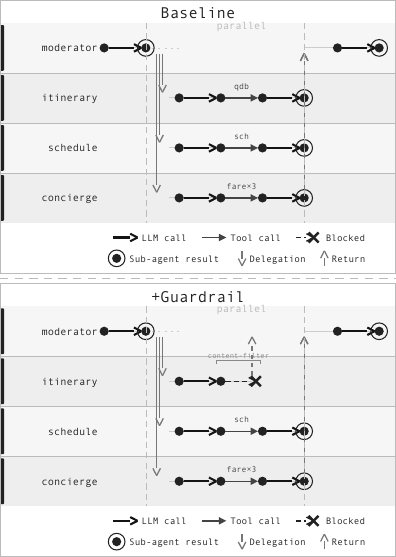}
  \caption{Trajectories for a fault-injected query (forbidden destination).
    Left: baseline.  Right: with the content filter active.
    Rows show per-agent steps; identical steps are highlighted.
    Three agents are completely unaffected.
    The itinerary agent shares its first step (LLM decision) then
    diverges: the call to the database is blocked before it executes.}
  \label{fig:lab2traj}
\end{figure}

\begin{figure}[htbp]
  \centering
  \includegraphics[width=\columnwidth]{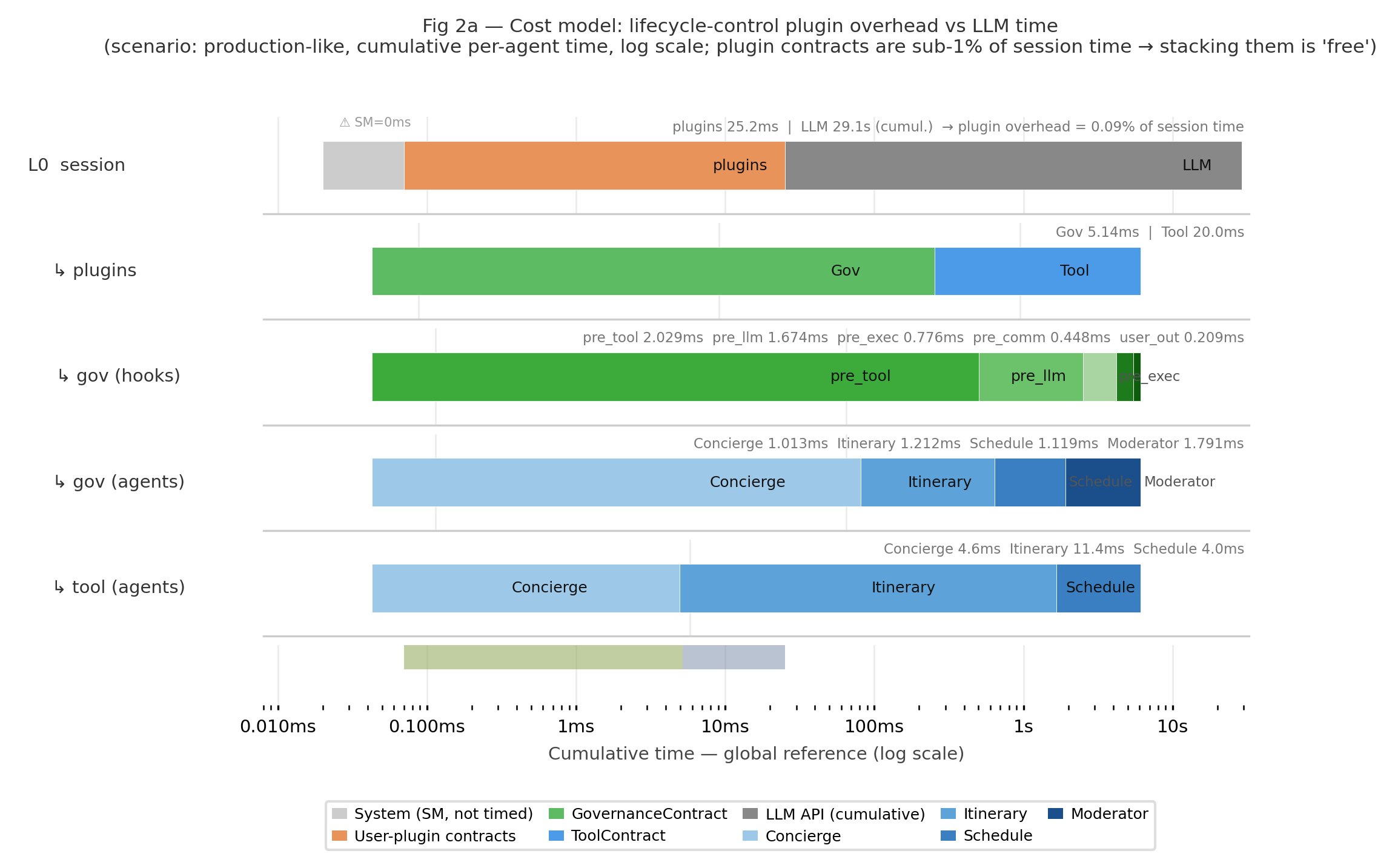}
  \caption{Time breakdown for a nominal query under the full production
    stack (log scale).
    LLM processing accounts for more than 99.8\% of total time.
    All governance evaluations combined contribute 0.12\%.}
  \label{fig:lab2cost}
\end{figure}

\begin{figure}[htbp]
  \centering
  \includegraphics[width=\columnwidth]{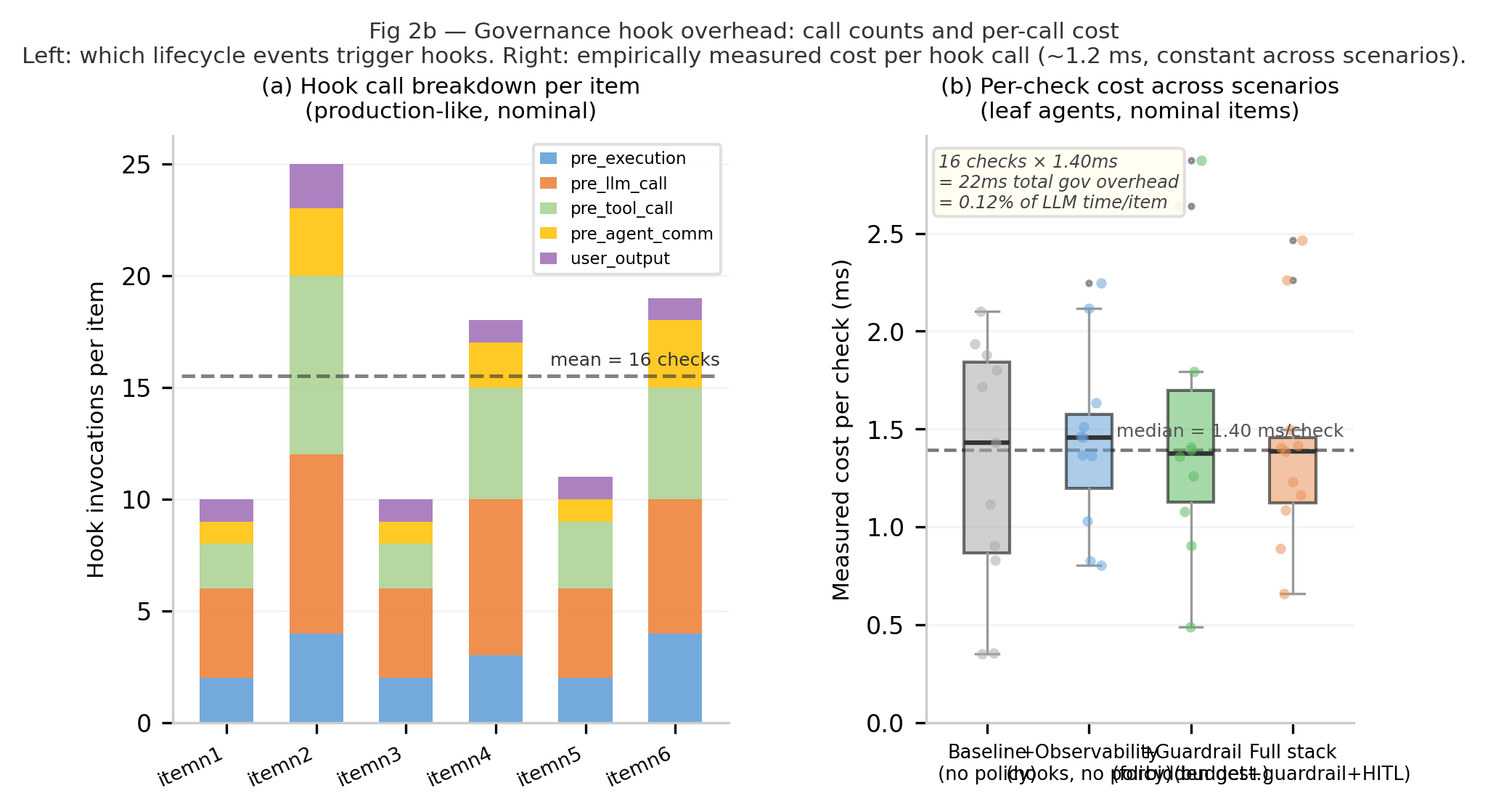}
  \caption{Governance evaluation count and cost per query (nominal items,
    full production stack).
    Left: how many times each type of check fires per query on average.
    Right: how long one check takes; stable at around 1.4\,ms across all
    scenarios.}
  \label{fig:lab2overhead}
\end{figure}

On the six nominal queries the full stack fires zero interventions;
the execution trace is identical to the baseline.
Fig.~\ref{fig:lab2traj} shows what happens on the fault-injected query:
three of four agents execute the same steps in both conditions.
For the itinerary agent the first step --- an LLM decision --- is shared
(the response cache ensures the model makes the same choice); at the next
step the content filter evaluates the tool argument, finds the blocked
destination, and denies the call.  The tool never runs.
The filter fires exactly once per matching attempt, and the circuit breaker
records a halt event after the declared number of consecutive tool failures.

Fig.~\ref{fig:lab2cost} shows that all governance evaluations together
account for 0.12\% of query time; the language model dominates entirely.
Fig.~\ref{fig:lab2overhead} breaks this down: observability checks are the
most numerous because they fire on every event.
The cost per check is stable at around 1.4\,ms; total overhead grows
with event count, not with policy complexity.
If a plugin introduced unexpected latency --- for instance by calling an
external service on every LLM invocation --- the spike would appear
immediately in the per-check breakdown without any added instrumentation.

Two properties of MAS-Lab made this measurement design possible.
First, the response cache is an infra-layer component declared in the
overlay, not a proxy inserted externally.%
\footnote{The MAS-OS state machine exposes execution checkpoints
  that can snapshot and replay from any intermediate state.  We did not
  use them here because the response cache already ensures identical LLM
  decisions.  Checkpoints would additionally allow injecting a fault at
  step $k$ of an otherwise normal run; this is left to future work.}
Second, because each plugin evaluates at a named point in a shared
execution trace, the cost attribution in Fig.~\ref{fig:lab2overhead}
is always present --- governance overhead is never a black box.
The deeper reason controls compose without side-effects is structural:
MAS-OS models execution as a product of independent state machines.
Each governance plugin, each observability subscriber, and the agent
are separate machines that share only the event stream.
Adding a new machine never modifies the transition logic of an existing one.
This is what makes the verification tractable: each control can be
tested in isolation, and their composition does not require a combined
testing surface.

Labs~1 and~2 have treated the agent's reasoning and control layer as the
primary variables.  Lab~3 shifts focus to the third dimension of the
lifecycle: \emph{what does the agent know?}  Context is assembled dynamically
from external sources at each turn; the choice of source --- a vector store,
a hosted memory service, a graph database --- affects recall, precision, cost,
and latency in ways that are hard to attribute without a unified
instrumentation layer.


\subsection{Lab~3: Context Source Extensions}
\label{sec:lab3}

Different memory libraries --- vector stores, hosted memory services, graph
databases --- have distinct retrieval semantics and no common attribution
trace.  Comparing them typically requires separate instrumentation pipelines
and manual reconciliation.  Lab~3 shows that wrapping any such library into
the framework's plugin contracts eliminates that overhead and makes the
comparison direct.

We use Letta~\cite{packer2024memgptllmsoperatingsystems}, an established memory library, as the primary
example.  Letta exposes three memory mechanisms: \emph{core memory}
(a fixed-size block always placed in the system prompt before the first LLM
call), \emph{recall memory} (search over past conversations, issued as a
tool call by the model), and \emph{archival memory} (long-term persistent
store, also accessed via tool call).  In this experiment we connect only
core memory.  The adapter is a single overlay file; no agent code changes.
For comparison we built a plain vector-similarity plugin as a reference
implementation --- here retrieval happens when the model decides to issue a
search, and the top-$k$ matching chunks are appended to the context.
Table~\ref{tab:lab3contracts} shows how each mechanism maps to a framework
contract; the experiment uses one from each column.

\begin{table}[htbp]
\caption{Letta memory mechanisms and how they map to MAS-Lab contracts.
  The experiment uses only core memory and the vector baseline; the other
  rows illustrate how a full Letta integration would be expressed.}
\label{tab:lab3contracts}
\small
\begin{tabular}{p{2.8cm}p{2.4cm}p{2.0cm}}
\toprule
\textbf{Memory mechanism} & \textbf{How it reaches the model} & \textbf{Who triggers it} \\
\midrule
Core memory (Letta)        & Pre-turn injection       & Framework  \\
Recall memory (Letta)      & Tool call during turn    & Model      \\
Archival memory (Letta)    & Tool call during turn    & Model      \\
Vector similarity (ours)   & Tool call during turn    & Model      \\
\bottomrule
\end{tabular}
\end{table}

Once a source is wrapped, every memory access produces a provenance record
in the shared trace, regardless of which library it came from.
Figure~\ref{fig:exp33-prompt} shows the two records side by side for a
representative query.  The Letta-core record shows a single pre-turn
injection event with all 100 facts always available; the RAG-direct record
shows a model-issued search returning 3 chunks.  The fields that capture how
the content was assembled and who triggered it are natively present in both
records without any extra instrumentation.  The \emph{key} field in each
record matches the identifier in the dataset ground truth, so recall and
precision are computed by set comparison --- no labelling required.

\begin{figure}[htbp]
\centering
\setlength{\tabcolsep}{4pt}
\renewcommand{\arraystretch}{1.18}
\footnotesize
\begin{minipage}[t]{0.46\columnwidth}
\centering
{\small\textbf{Letta-core}}\\
{\scriptsize\textit{pre-turn injection}}
\vspace{4pt}
\begin{tabular}{@{} p{2.1cm} p{1.55cm} @{}}
\toprule
\multicolumn{2}{@{}l}{\textit{before first LLM call}} \\
\midrule
role instructions & {\textasciitilde}152 tok. \\
available agents  & {\textasciitilde}87 tok. \\
\rowcolor{diffrow}
memory block      & 1{,}881 tok. \\
\rowcolor{diffrow}
                  & (100 facts) \\
\midrule
\multicolumn{2}{@{}l}{\textit{during turn}} \\
\midrule
\rowcolor{diffrow}
memory search     & \textit{not issued} \\
\midrule
\multicolumn{2}{@{}l}{\textit{provenance record}} \\
\midrule
\rowcolor{diffrow}
assembled by      & framework \\
\rowcolor{diffrow}
triggered by      & deterministic \\
\rowcolor{diffrow}
keys in trace     & f001\,\ldots\,f100 \\
\rowcolor{samerow}
ground truth      & f004\,\checkmark \\
\bottomrule
\end{tabular}
\end{minipage}
\hfill
\begin{minipage}[t]{0.46\columnwidth}
\centering
{\small\textbf{RAG-direct}}\\
{\scriptsize\textit{tool-call retrieval}}
\vspace{4pt}
\begin{tabular}{@{} p{2.1cm} p{1.55cm} @{}}
\toprule
\multicolumn{2}{@{}l}{\textit{before first LLM call}} \\
\midrule
role instructions & {\textasciitilde}137 tok. \\
available agents  & {\textasciitilde}87 tok. \\
\rowcolor{absentrow}
memory block      & \textit{absent} \\
\midrule
\multicolumn{2}{@{}l}{\textit{during turn}} \\
\midrule
\rowcolor{diffrow}
memory search     & 3 chunks \\
\rowcolor{diffrow}
                  & 129 tok. \\
\midrule
\multicolumn{2}{@{}l}{\textit{provenance record}} \\
\midrule
\rowcolor{diffrow}
assembled by      & model \\
\rowcolor{diffrow}
triggered by      & stochastic \\
\rowcolor{diffrow}
keys in trace     & f004, f056, f075 \\
\rowcolor{samerow}
ground truth      & f004\,\checkmark \\
\bottomrule
\end{tabular}
\end{minipage}
\vspace{4pt}
\begin{flushleft}
\scriptsize
\colorbox{diffrow}{\phantom{x}}~fields that differ between modes\quad
\colorbox{samerow}{\phantom{x}}~fields shared by all provenance records\quad
\colorbox{absentrow}{\phantom{x}}~segment absent in this mode
\end{flushleft}
\caption{Lab~3: prompt structure and provenance record for the same query
  (\emph{k1\_q04}: ``What is my daily travel budget?'')
  under two memory integrations.  Yellow rows are fields that differ;
  grey rows are common to every record.
  The \emph{keys in trace} field links to the ground-truth identifier (f004),
  enabling recall computation from the trace alone.}
\label{fig:exp33-prompt}
\end{figure}

\subsubsection{Exp~3.3: Comparing Context Sources}
\label{sec:exp33}

Four conditions share the same agent, the same 100-fact personal profile for
traveler Alex (\texttt{f001}--\texttt{f100}), and the same 10-item dataset.
Items are grouped by how many personal facts the correct answer requires:
none (k0), one (k1), two (k2), or three (k3).  k0 items require no
personal facts and all conditions answer them correctly; figures show
only k1--k3, where the memory strategy matters.

\begin{itemize}
\item \textbf{Letta-core}: Letta core memory, pre-turn injection.  All 100
  facts always present; recall~$= 1.0$ by construction.
\item \textbf{RAG-direct (k=6)}: vector similarity, single search per
  query, top-6 results.
\item \textbf{RAG-decomposed (k=6)}: same vector store, agent instructed
  to issue multiple targeted searches for multi-fact questions.
\item \textbf{RAG-wide (k=20)}: same vector store, single search, top-20
  results; broader recall at the cost of lower precision.
\end{itemize}

Letta-core establishes an analytical upper bound on recall; the price is
that all 100 facts are always injected regardless of relevance, which is
visible in precision scores near the floor.  The three vector conditions
explore the precision--recall space achievable with selective retrieval.

\begin{figure}[htbp]
\centering
\begin{subfigure}[t]{\columnwidth}
  \includegraphics[width=\columnwidth]{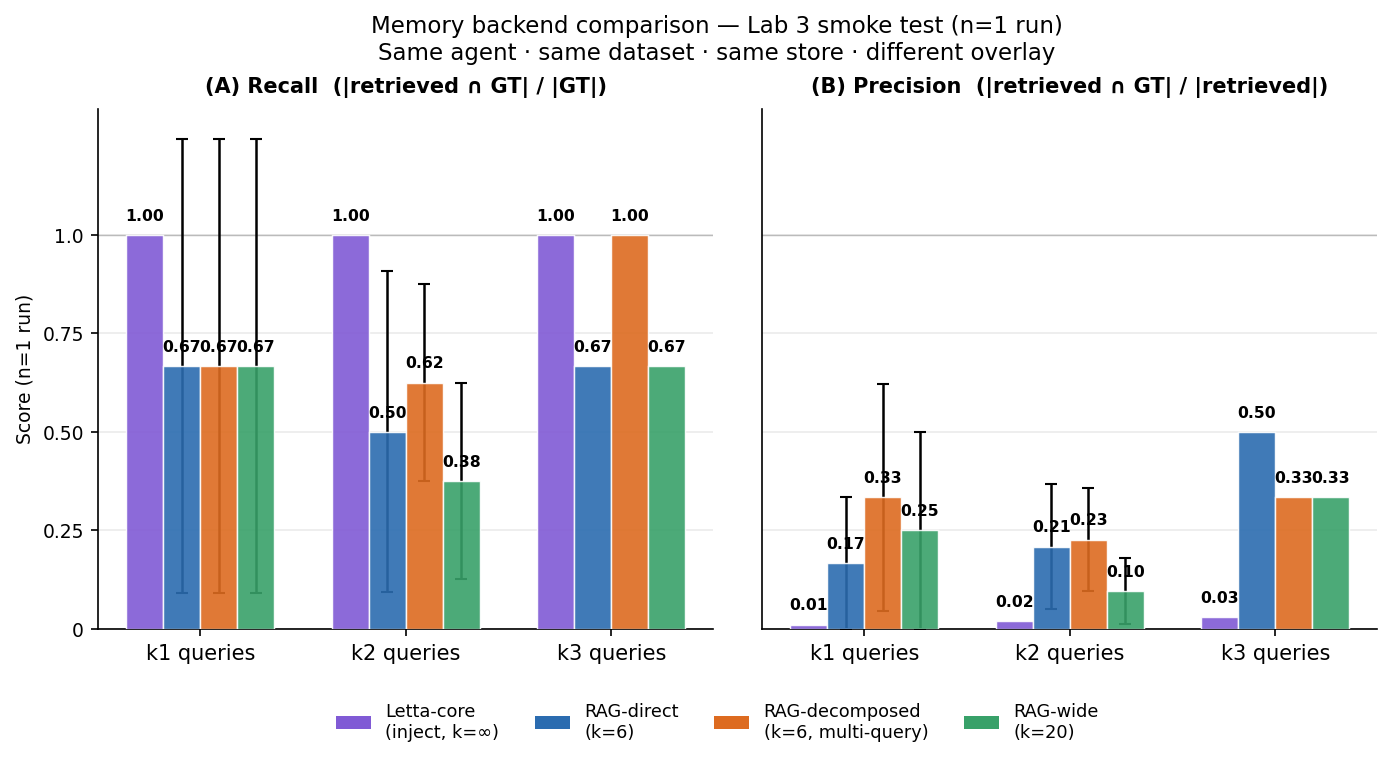}
  \caption{Recall and precision by condition and query group ($n{=}1$).
    Error bars show score dispersion across items within each group;
    per-run confidence intervals require multiple runs.}
  \label{fig:exp33-recall}
\end{subfigure}
\vspace{0.8em}
\begin{subfigure}[t]{\columnwidth}
  \includegraphics[width=\columnwidth]{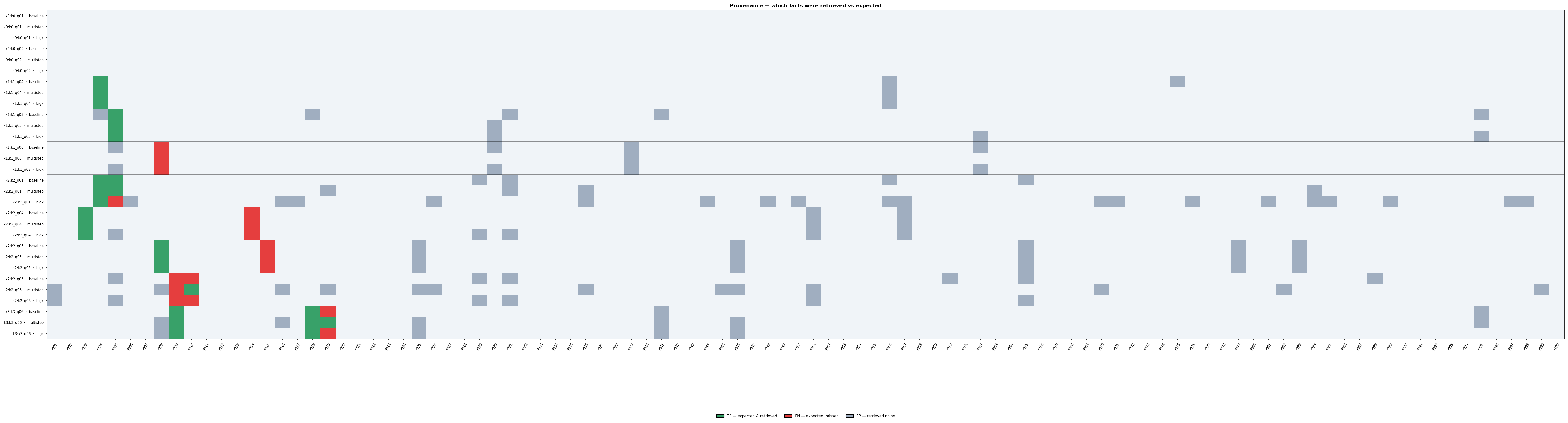}
  \caption{Fact-level attribution: TP (green), FN (red), FP (grey) per
    (item, condition) row.  Ground-truth labels come from the dataset;
    retrieved labels come from the \emph{key} field in each provenance
    record.  No human annotation is required.}
  \label{fig:exp33-provenance}
\end{subfigure}
\caption{Lab~3, Exp~3.3: recall--precision and fact-level attribution
  (4 conditions, 10-item smoke run, $n{=}1$).}
\label{fig:exp33}
\end{figure}

\begin{figure}[htbp]
\centering
\includegraphics[width=0.75\columnwidth]{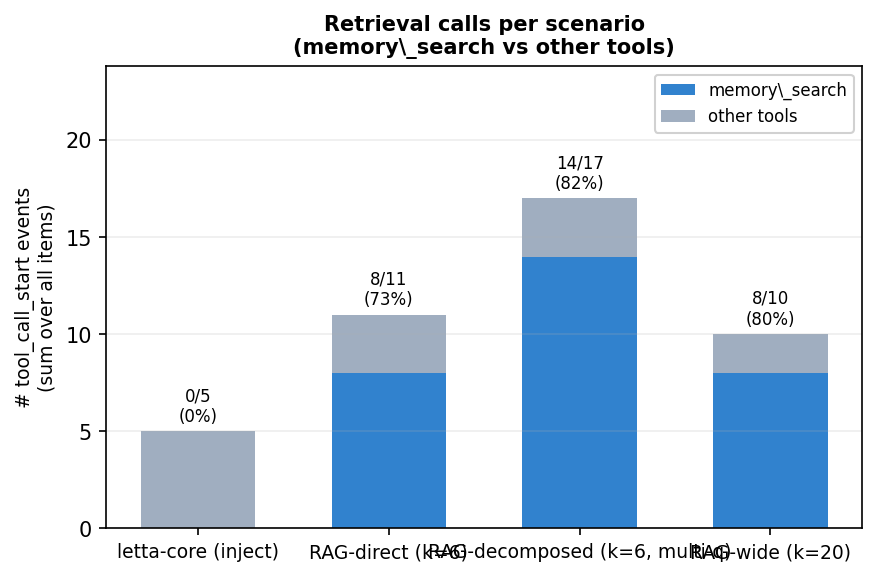}
\caption{Lab~3, Exp~3.3: memory searches vs.\ other tool calls per
  condition.  Letta-core issues no searches (content injected before each
  turn).  RAG-decomposed issues the most, confirming the multi-query
  instruction was followed.  RAG-wide makes fewer other-tool calls than
  RAG-direct; MAS-Lab's per-session tool-call attribution makes such
  hypotheses directly testable across conditions without additional
  instrumentation.}
\label{fig:exp33-calls}
\end{figure}

On k3\_q06 --- requiring three semantically independent facts (loyalty
programme, travel insurance, payment card) --- RAG-decomposed achieves
$R{=}1.0$ while RAG-direct and RAG-wide reach only $R{=}0.67$.  A single
combined query pulls two of the three facts to the top and crowds out the
third; three targeted sub-queries each surface their target independently.
On k2\_q06 --- requiring a loyalty number and a seat preference, two
cross-domain facts --- RAG-direct and RAG-wide both return $R{=}0.0$:
neither fact anchors the other in the embedding space, so the combined
query returns neither.  RAG-decomposed recovers one of the two ($R{=}0.5$).
Letta-core achieves $R{=}1.0$ on every item by construction.

The fact-level heatmap (Figure~\ref{fig:exp33-provenance}) makes the failure
mode visible: a red cell is a fact the dataset expects but the trace does not
contain; a grey cell is a fact the trace contains but the dataset does not
expect.  This distinction --- which aggregate recall cannot make ---
is directly readable from the provenance records without any annotation.
Adding Letta to the comparison required one overlay file; the agent, dataset,
and pipeline were unchanged.


\section{Discussion}
\label{sec:discussion}

The three labs exercise successive stages of the agent lifecycle and
collectively show that design exploration, operational control, and context
attribution can share a single declarative specification layer and a single
trace schema.  A design chosen in Lab~1 runs under the governance stack from
Lab~2, and its context accesses are traced with the provenance schema from
Lab~3.  The experiments cover a limited surface deliberately: each lab
isolates one question so the measurement is unambiguous.  What they validate
is infrastructure, not a claim about which design choices are universally best.

\paragraph{Protocol alignment and interoperability.}
A recurring design question for MAS infrastructure is where agent-to-agent
and agent-to-tool communication protocols fit.  The contract interfaces
described in this paper --- tool contracts, context contracts, governance
hooks, delegation edges --- are semantically very close to what emerging
protocols such as MCP~\cite{anthropic_mcp_2024} and A2A~\cite{a2a} formalise at the wire
level.  The relationship is one of mutual reinforcement rather than
competition: the declarative contracts in MAS-Lab provide a local,
testable semantics for what MCP and A2A encode as protocol messages, and
a production-grade MCP or A2A implementation can be connected as a plugin
without changing the agent specification or the evaluation pipeline.%
\footnote{A full MCP and A2A integration, including the corresponding plugin
  adapters and conformance tests, is planned for a forthcoming open-source
  release.  The contract decoupling means the integration can be developed
  and shipped independently of agent logic, and its correctness can be
  tested by running the same experiment specifications with the protocol layer
  substituted in.}
This alignment has a practical implication for adoption: any framework
that already exposes MCP or A2A endpoints can participate in a MAS-Lab
evaluation by wrapping those endpoints as contracts, without re-implementing
agent logic.  The framework provides the measurement infrastructure; the
framework under evaluation provides the behaviour.

\paragraph{Runtime architecture and local testability.}
The runtime core is implemented in Python; thanks to the contract interfaces
decoupling the core from the plugins, it could be reimplemented in
Rust, giving a stable, memory-safe substrate for the state machine, event bus,
and plugin registry.  Plugins that handle sensitive operations --- budget
enforcement, content filtering, external tool calls --- can be isolated in
separate OS processes and communicate with the core via the egress protocol,
providing fault isolation and a security boundary that prevents a misbehaving
plugin from accessing agent state it should not see.  A practical consequence
of this architecture is that local testing does not require network services.
The same experiment specification that in production routes delegation through an
A2A-compatible transport runs locally with an in-process bus --- same agents,
same overlays, same trace schema.  This two-level testing pattern --- logic
first with the local bus, then protocol validation with the real transport ---
is a direct application of the decoupling principle: the correct way to test
whether an A2A integration adds the expected behaviour is to hold the agent
logic constant and vary only the transport layer.  Massive parallel regression
suites run at the logic level; protocol-layer tests are a smaller, targeted
addition.

\paragraph{Reference basis and reusability.}
Each lab ships as a self-contained artifact: experiment specification, dataset,
overlay set, and analysis pipeline are versioned together and produce a
reproducible PDF from a single command.  This is intentional.  A team
adopting the framework can fork a published experiment, substitute their
own application, and immediately know whether their agent matches the
reference on the same metrics --- without writing evaluation code.  The
same principle applies at the framework level: MAS-Lab does not try to be
the only way to build agents; it tries to be a reliable reference point
that any toolchain can measure against by implementing the contract
interfaces.

\paragraph{Reproducibility.}
Each experiment is fully specified in a versioned YAML specification committed
alongside its overlays, datasets, and pipeline.  On re-run, the same inputs
produce the same trace events; LLM stochasticity is mitigated by structural
(not textual) assertions and multiple runs.  Result metadata includes model
name, date, MAS-Lab version, dataset version, and commit hash.  A team can
fork a specification, update the model endpoint, and re-run the full analysis
pipeline with one command; the experiment is the unit of scientific record.

\paragraph{Moving to production.}
Each lab currently executes in a local deployment flavor; however, transitioning to production requires only declarative specification changes, such as selecting an A2A-based plugin that implements the TransportContract. The current plugin ecosystem is primarily focused on local execution, but we plan to extend it with production-grade backends (i.e. agents deployed as docker containers) and additional agent-to-agent communication protocols. Importantly, \textit{MAS-Lab}’s architecture is intentionally extensible: new plugins can be introduced without modifying the core runtime, while preserving the existing contract interfaces.

\section{Conclusions}
\label{sec:conclusions}

We presented \textit{MAS-Lab}, a specification-driven framework for building, validating, and operating multi-agent systems across the full development lifecycle.
The central argument is not that MAS are better than single agents in general, nor that specific reasoning patterns outperform others.
The central argument is that \emph{the class of questions a developer can answer about their MAS depends on the infrastructure underneath it.}
Today, key design questions---does adding a memory plugin improve quality? is this topology worth the coordination overhead? does a governance control fire in production as it did in staging?---are either answered by intuition, or require bespoke per-project evaluation pipelines that are not reusable or reproducible.

\textit{MAS-Lab} changes this by making three properties structural rather than accidental.
\emph{Declarative-first}: every aspect of a MAS that can affect its behaviour---agent roles, tool access, reasoning patterns, governance controls---is expressed in versioned YAML before any code runs.
This makes design dimensions into first-class variables, comparable across conditions in a single experiment specification, without modifying agent logic.
\emph{Boundaries, not internals}: contracts constrain only what crosses I/O boundaries, leaving agent reasoning unconstrained while making every context injection, tool invocation, and delegation step visible and auditable.
\emph{Observable by construction}: every boundary crossing produces a trace event, and every quality claim produced by the lab layer is linked to that event stream.
Quality differences are \emph{explained}, not asserted.

The experimental section provides concrete evidence for these properties across three labs:
Lab~1 demonstrates that five reasoning patterns and five topologies can be compared by YAML swap alone;
Lab~2 demonstrates that governance controls can be added, verified, and promoted to production as conformance regression gates without agent code modification;
Lab~3 demonstrates that memory strategy and topology choices produce trace-attributable quality differences---including null results that quantify when coordination overhead is unjustified.
More broadly, the separation of topology from routing and forwarding decisions
opens the contract interface to third-party coordination mechanisms---Paxos-style
consensus, task allocation policies, semantic dispatching routers---making
MAS-Lab a substrate for studying the cognitive science of agent collaboration
beyond the infrastructure questions addressed here.

\textit{MAS-Lab} is open-source and includes progressive tutorials that exercise each layer of the stack, from a single agent specification to a multi-agent topology comparison with automated evaluation pipelines.
The experiment specifications and datasets for all three labs are released as reproducible artifacts; forking a specification and changing one parameter is sufficient to extend any experiment.

\bibliographystyle{ACM-Reference-Format}
\bibliography{MAS-Lab}


\begin{thebibliography}{37}


\ifx \showCODEN    \undefined \def \showCODEN     #1{\unskip}     \fi
\ifx \showISBNx    \undefined \def \showISBNx     #1{\unskip}     \fi
\ifx \showISBNxiii \undefined \def \showISBNxiii  #1{\unskip}     \fi
\ifx \showISSN     \undefined \def \showISSN      #1{\unskip}     \fi
\ifx \showLCCN     \undefined \def \showLCCN      #1{\unskip}     \fi
\ifx \shownote     \undefined \def \shownote      #1{#1}          \fi
\ifx \showarticletitle \undefined \def \showarticletitle #1{#1}   \fi
\ifx \showURL      \undefined \def \showURL       {\relax}        \fi
\providecommand\bibfield[2]{#2}
\providecommand\bibinfo[2]{#2}
\providecommand\natexlab[1]{#1}
\providecommand\showeprint[2][]{arXiv:#2}

\bibitem[mcp(2025)]%
        {mcp_spec_2025}
 \bibinfo{year}{2025}\natexlab{}.
\newblock \bibinfo{title}{Model Context Protocol (MCP) Specification}.
\newblock
  \bibinfo{howpublished}{\url{https://modelcontextprotocol.io/specification/2025-06-18}}.
\newblock


\bibitem[ope(2025)]%
        {opentelemetry_docs_2025}
 \bibinfo{year}{2025}\natexlab{}.
\newblock \bibinfo{title}{OpenTelemetry Documentation}.
\newblock \bibinfo{howpublished}{\url{https://opentelemetry.io/docs/}}.
\newblock


\bibitem[lan(2026a)]%
        {langchain_docs_2026}
 \bibinfo{year}{2026}\natexlab{a}.
\newblock \bibinfo{title}{LangChain Documentation}.
\newblock \bibinfo{howpublished}{\url{https://docs.langchain.com/}}.
\newblock


\bibitem[lan(2026b)]%
        {langsmith_observability_2026}
 \bibinfo{year}{2026}\natexlab{b}.
\newblock \bibinfo{title}{LangSmith: AI Agent \& LLM Observability Platform}.
\newblock
  \bibinfo{howpublished}{\url{https://www.langchain.com/langsmith/observability}}.
\newblock


\bibitem[rag(2026)]%
        {ragas_docs_2026}
 \bibinfo{year}{2026}\natexlab{}.
\newblock \bibinfo{title}{Ragas Documentation: Evaluate LLM Applications}.
\newblock
  \bibinfo{howpublished}{\url{https://docs.ragas.io/en/stable/getstarted/evals/}}.
\newblock


\bibitem[tru(2026)]%
        {trulens_2026}
 \bibinfo{year}{2026}\natexlab{}.
\newblock \bibinfo{title}{TruLens: Evals and Tracing for Agents}.
\newblock \bibinfo{howpublished}{\url{https://www.trulens.org/}}.
\newblock


\bibitem[{AGNTCY Project}(2025)]%
        {telemetryhub2025}
\bibfield{author}{\bibinfo{person}{{AGNTCY Project}}.}
  \bibinfo{year}{2025}\natexlab{}.
\newblock \bibinfo{title}{Metrics Computation Engine ({MCE}): Metrics from
  {OTel} Observability Telemetry}.
\newblock
  \bibinfo{howpublished}{\url{https://github.com/agntcy/telemetry-hub}}.
\newblock


\bibitem[{AGNTCY Project}(2026)]%
        {oasf}
\bibfield{author}{\bibinfo{person}{{AGNTCY Project}}.}
  \bibinfo{year}{2026}\natexlab{}.
\newblock \bibinfo{title}{Open Agentic Schema Framework (OASF)}.
\newblock
  \bibinfo{howpublished}{\url{https://docs.agntcy.org/oasf/open-agentic-schema-framework/}}.
\newblock


\bibitem[Amini et~al\mbox{.}(2025)]%
        {agent_spec_arxiv_2025}
\bibfield{author}{\bibinfo{person}{Soufiane Amini}, \bibinfo{person}{Yassine
  Benajiba}, \bibinfo{person}{Cesare Bernardis}, \bibinfo{person}{Paul Cayet},
  \bibinfo{person}{Hassan Chafi}, \bibinfo{person}{Abderrahim Fathan},
  \bibinfo{person}{Louis Faucon}, \bibinfo{person}{Damien Hilloulin},
  \bibinfo{person}{Sungpack Hong}, \bibinfo{person}{Ingo Kossyk},
  \bibinfo{person}{Tran Minh~Son Le}, \bibinfo{person}{Rhicheek Patra},
  \bibinfo{person}{Sujith Ravi}, \bibinfo{person}{Jonas Schweizer},
  \bibinfo{person}{Jyotika Singh}, \bibinfo{person}{Shailender Singh},
  \bibinfo{person}{Weiyi Sun}, \bibinfo{person}{Kartik Talamadupula}, {and}
  \bibinfo{person}{Jerry Xu}.} \bibinfo{year}{2025}\natexlab{}.
\newblock \showarticletitle{Open Agent Specification (Agent Spec): A Unified
  Representation for AI Agents}.
\newblock \bibinfo{journal}{\emph{arXiv}} (\bibinfo{year}{2025}).
\newblock
\showeprint{2510.04173}
\urldef\tempurl%
\url{https://arxiv.org/abs/2510.04173}
\showURL{%
\tempurl}


\bibitem[{Anthropic}(2024a)]%
        {anthropic_mcp_2024}
\bibfield{author}{\bibinfo{person}{{Anthropic}}.}
  \bibinfo{year}{2024}\natexlab{a}.
\newblock \bibinfo{title}{Introducing the Model Context Protocol}.
\newblock
  \bibinfo{howpublished}{\url{https://www.anthropic.com/news/model-context-protocol}}.
\newblock


\bibitem[{Anthropic}(2024b)]%
        {anthropic_tools}
\bibfield{author}{\bibinfo{person}{{Anthropic}}.}
  \bibinfo{year}{2024}\natexlab{b}.
\newblock \bibinfo{title}{Tool use with Claude}.
\newblock
  \bibinfo{howpublished}{\url{https://docs.anthropic.com/en/docs/build-with-claude/tool-use}}.
\newblock


\bibitem[Capucci(2026)]%
        {agentmanifest}
\bibfield{author}{\bibinfo{person}{Hern{\'a}n~Alfredo Capucci}.}
  \bibinfo{year}{2026}\natexlab{}.
\newblock \bibinfo{title}{Agent Manifest: Core Declarative Specification v1.0}.
\newblock
  \bibinfo{howpublished}{\url{https://agent-manifest-spec.org/spec/v1.0/agent_manifest_v1.0.html}}.
\newblock


\bibitem[Casel(2025)]%
        {agentos_buildermethods}
\bibfield{author}{\bibinfo{person}{Brian Casel}.}
  \bibinfo{year}{2025}\natexlab{}.
\newblock \bibinfo{title}{{Agent OS}: A System for Spec-Driven Development with
  AI Agents}.
\newblock
  \bibinfo{howpublished}{\url{https://github.com/buildermethods/agent-os}}.
\newblock
\newblock
\shownote{Open-source project, accessed April 2026}.


\bibitem[{Confident AI}(2024)]%
        {deepeval}
\bibfield{author}{\bibinfo{person}{{Confident AI}}.}
  \bibinfo{year}{2024}\natexlab{}.
\newblock \bibinfo{title}{{DeepEval}: The {LLM} Evaluation Framework}.
\newblock
  \bibinfo{howpublished}{\url{https://github.com/confident-ai/deepeval}}.
\newblock


\bibitem[{CrewAI Contributors}(2024)]%
        {crewai_2024}
\bibfield{author}{\bibinfo{person}{{CrewAI Contributors}}.}
  \bibinfo{year}{2024}\natexlab{}.
\newblock \bibinfo{title}{CrewAI: A Framework for Orchestrating Role-Based AI
  Agents}.
\newblock \bibinfo{howpublished}{\url{https://github.com/joaomdmoura/crewai}}.
\newblock


\bibitem[{deepset}(2024)]%
        {haystack_agents_2024}
\bibfield{author}{\bibinfo{person}{{deepset}}.}
  \bibinfo{year}{2024}\natexlab{}.
\newblock \bibinfo{title}{Haystack Agents}.
\newblock \bibinfo{howpublished}{\url{https://haystack.deepset.ai}}.
\newblock


\bibitem[{Google and Industry Contributors}(2025)]%
        {a2a}
\bibfield{author}{\bibinfo{person}{{Google and Industry Contributors}}.}
  \bibinfo{year}{2025}\natexlab{}.
\newblock \bibinfo{title}{Agent2Agent (A2A) Protocol}.
\newblock \bibinfo{howpublished}{\url{https://a2a-protocol.org}}.
\newblock
\newblock
\shownote{Emerging standard for agent interoperability; see also
  \url{https://github.com/a2aproject/A2A}}.


\bibitem[{Google DeepMind \& Google Cloud}(2026)]%
        {google_adk_2026}
\bibfield{author}{\bibinfo{person}{{Google DeepMind \& Google Cloud}}.}
  \bibinfo{year}{2026}\natexlab{}.
\newblock \bibinfo{title}{Google Agent Development Kit (ADK)}.
\newblock \bibinfo{howpublished}{\url{https://google.github.io/adk-docs/}}.
\newblock


\bibitem[Gosmar and Dahl(2025)]%
        {sentinel_agents_2025}
\bibfield{author}{\bibinfo{person}{Diego Gosmar} {and}
  \bibinfo{person}{Deborah~A. Dahl}.} \bibinfo{year}{2025}\natexlab{}.
\newblock \showarticletitle{Sentinel Agents for Secure and Trustworthy Agentic
  AI in Multi-Agent Systems}.
\newblock \bibinfo{journal}{\emph{arXiv}} (\bibinfo{year}{2025}).
\newblock
\showeprint{2509.14956}
\urldef\tempurl%
\url{https://arxiv.org/abs/2509.14956}
\showURL{%
\tempurl}


\bibitem[H{\"a}rer(2025)]%
        {mas_schema}
\bibfield{author}{\bibinfo{person}{Felix H{\"a}rer}.}
  \bibinfo{year}{2025}\natexlab{}.
\newblock \showarticletitle{Specification and Evaluation of Multi-Agent LLM
  Systems: Prototype and Cybersecurity Applications}.
\newblock \bibinfo{journal}{\emph{arXiv preprint arXiv:2506.10467}}
  (\bibinfo{year}{2025}).
\newblock


\bibitem[Khattab et~al\mbox{.}(2024)]%
        {dspy_2024}
\bibfield{author}{\bibinfo{person}{Omar Khattab}, \bibinfo{person}{Arnav
  Singhvi}, \bibinfo{person}{Paridhi Maheshwari}, \bibinfo{person}{Zhiyuan
  Zhang}, \bibinfo{person}{Keshav Santhanam}, \bibinfo{person}{Sri
  Vardhamanan}, \bibinfo{person}{Saiful Haq}, \bibinfo{person}{Ashutosh
  Sharma}, \bibinfo{person}{Thomas~T. Joshi}, \bibinfo{person}{Hanna Moazam},
  \bibinfo{person}{Heather Miller}, \bibinfo{person}{Matei Zaharia}, {and}
  \bibinfo{person}{Christopher Potts}.} \bibinfo{year}{2024}\natexlab{}.
\newblock \showarticletitle{DSPy: Compiling Declarative Language Model Calls
  into Self-Improving Pipelines}.
\newblock \bibinfo{journal}{\emph{arXiv preprint arXiv:2310.03714}}
  (\bibinfo{year}{2024}).
\newblock


\bibitem[{LangChain}(2026)]%
        {langgraph_overview_2026}
\bibfield{author}{\bibinfo{person}{{LangChain}}.}
  \bibinfo{year}{2026}\natexlab{}.
\newblock \bibinfo{title}{LangGraph Overview — LangChain Docs (Python)}.
\newblock \bibinfo{howpublished}{Online documentation}.
\newblock
\urldef\tempurl%
\url{https://docs.langchain.com/oss/python/langgraph/overview}
\showURL{%
\tempurl}


\bibitem[Liang et~al\mbox{.}(2022)]%
        {helm_eval}
\bibfield{author}{\bibinfo{person}{Percy Liang}, \bibinfo{person}{Rishi
  Bommasani}, \bibinfo{person}{Tony Lee}, {et~al\mbox{.}}}
  \bibinfo{year}{2022}\natexlab{}.
\newblock \showarticletitle{Holistic Evaluation of Language Models}.
\newblock \bibinfo{journal}{\emph{arXiv preprint arXiv:2211.09110}}
  (\bibinfo{year}{2022}).
\newblock


\bibitem[{Linux Foundation AI \& Data}(2026)]%
        {agntcy}
\bibfield{author}{\bibinfo{person}{{Linux Foundation AI \& Data}}.}
  \bibinfo{year}{2026}\natexlab{}.
\newblock \bibinfo{title}{AGNTCY: Open Infrastructure for Agent
  Interoperability}.
\newblock \bibinfo{howpublished}{\url{https://agntcy.org}}.
\newblock


\bibitem[Liu et~al\mbox{.}(2024)]%
        {agentbench_liu_2024}
\bibfield{author}{\bibinfo{person}{Xiao Liu}, \bibinfo{person}{Hao Yu},
  \bibinfo{person}{Hanchen Zhang}, {et~al\mbox{.}}}
  \bibinfo{year}{2024}\natexlab{}.
\newblock \showarticletitle{AgentBench: Evaluating LLMs as Agents}.
\newblock \bibinfo{journal}{\emph{ICLR 2024}} (\bibinfo{year}{2024}).
\newblock
\urldef\tempurl%
\url{https://arxiv.org/abs/2308.03688}
\showURL{%
\tempurl}


\bibitem[{LlamaIndex}(2026)]%
        {llamaindex_2026}
\bibfield{author}{\bibinfo{person}{{LlamaIndex}}.}
  \bibinfo{year}{2026}\natexlab{}.
\newblock \bibinfo{title}{LlamaIndex Documentation}.
\newblock \bibinfo{howpublished}{\url{https://docs.llamaindex.ai/}}.
\newblock


\bibitem[Ma et~al\mbox{.}(2026)]%
        {maestro_2026}
\bibfield{author}{\bibinfo{person}{Tie Ma}, \bibinfo{person}{Yixi Chen},
  \bibinfo{person}{Vaastav Anand}, \bibinfo{person}{Alessandro Cornacchia},
  \bibinfo{person}{Am{\^a}ndio~R. Faustino}, \bibinfo{person}{Guanheng Liu},
  \bibinfo{person}{Shan Zhang}, \bibinfo{person}{Hongbin Luo},
  \bibinfo{person}{Suhaib~A. Fahmy}, \bibinfo{person}{Zafar~A. Qazi}, {and}
  \bibinfo{person}{Marco Canini}.} \bibinfo{year}{2026}\natexlab{}.
\newblock \showarticletitle{MAESTRO: Multi-Agent Evaluation Suite for Testing,
  Reliability, and Observability}.
\newblock \bibinfo{journal}{\emph{arXiv}} (\bibinfo{year}{2026}).
\newblock
\showeprint{2601.00481}
\urldef\tempurl%
\url{https://arxiv.org/abs/2601.00481}
\showURL{%
\tempurl}


\bibitem[{Microsoft Corporation}(2026)]%
        {microsoft_agent_framework_2026}
\bibfield{author}{\bibinfo{person}{{Microsoft Corporation}}.}
  \bibinfo{year}{2026}\natexlab{}.
\newblock \bibinfo{title}{Microsoft Agent Framework}.
\newblock
  \bibinfo{howpublished}{\url{https://github.com/microsoft/agent-framework}}.
\newblock


\bibitem[{Oracle Labs}(2025)]%
        {agent_spec_oracle_2025}
\bibfield{author}{\bibinfo{person}{{Oracle Labs}}.}
  \bibinfo{year}{2025}\natexlab{}.
\newblock \bibinfo{title}{Open Agent Specification (AgentSpec)}.
\newblock \bibinfo{howpublished}{\url{https://github.com/oracle/agent-spec}}.
\newblock


\bibitem[{Outshift Open}(2025)]%
        {outshiftopen2025}
\bibfield{author}{\bibinfo{person}{{Outshift Open}}.}
  \bibinfo{year}{2025}\natexlab{}.
\newblock \bibinfo{title}{{MAS-Lab}: Open-source Lab Implementations for
  Multi-Agent System Evaluation}.
\newblock
  \bibinfo{howpublished}{\url{https://github.com/outshift-open/mas-lab}}.
\newblock


\bibitem[Packer et~al\mbox{.}(2024)]%
        {packer2024memgptllmsoperatingsystems}
\bibfield{author}{\bibinfo{person}{Charles Packer}, \bibinfo{person}{Sarah
  Wooders}, \bibinfo{person}{Kevin Lin}, \bibinfo{person}{Vivian Fang},
  \bibinfo{person}{Shishir~G. Patil}, \bibinfo{person}{Ion Stoica}, {and}
  \bibinfo{person}{Joseph~E. Gonzalez}.} \bibinfo{year}{2024}\natexlab{}.
\newblock \bibinfo{title}{MemGPT: Towards LLMs as Operating Systems}.
\newblock
\showeprint[arxiv]{2310.08560}~[cs.AI]
\urldef\tempurl%
\url{https://arxiv.org/abs/2310.08560}
\showURL{%
\tempurl}


\bibitem[Qin et~al\mbox{.}(2023)]%
        {toolbench}
\bibfield{author}{\bibinfo{person}{Yujia Qin}, \bibinfo{person}{Shihao Liang},
  \bibinfo{person}{Yining Ye}, \bibinfo{person}{Kunlun Zhu},
  \bibinfo{person}{Lan Yan}, \bibinfo{person}{Yaxi Lu}, \bibinfo{person}{Yankai
  Lin}, \bibinfo{person}{Xin Cong}, \bibinfo{person}{Xiangru Tang},
  \bibinfo{person}{Bill Qian}, \bibinfo{person}{Sihan Zhao},
  \bibinfo{person}{Lauren Hong}, \bibinfo{person}{Runchu Tian},
  \bibinfo{person}{Ruobing Xie}, \bibinfo{person}{Jie Zhou},
  \bibinfo{person}{Mark Gerstein}, \bibinfo{person}{Dahai Li},
  \bibinfo{person}{Zhiyuan Liu}, {and} \bibinfo{person}{Maosong Sun}.}
  \bibinfo{year}{2023}\natexlab{}.
\newblock \showarticletitle{ToolLLM: Facilitating Large Language Models to
  Master 16000+ Real-world APIs}.
\newblock \bibinfo{journal}{\emph{arXiv preprint arXiv:2307.16789}}
  (\bibinfo{year}{2023}).
\newblock


\bibitem[Rivet(2026)]%
        {agentos_rivet}
\bibfield{author}{\bibinfo{person}{Rivet}.} \bibinfo{year}{2026}\natexlab{}.
\newblock \bibinfo{title}{agent-os: A Portable Open-Source Operating System for
  AI Agents}.
\newblock \bibinfo{howpublished}{\url{https://github.com/rivet-dev/agent-os}}.
\newblock
\newblock
\shownote{Open-source project, accessed April 2026}.


\bibitem[Wu et~al\mbox{.}(2024)]%
        {autogen_2024}
\bibfield{author}{\bibinfo{person}{Qingyun Wu} {et~al\mbox{.}}}
  \bibinfo{year}{2024}\natexlab{}.
\newblock \bibinfo{title}{AutoGen: Enabling Next-Gen LLM Applications via
  Multi-Agent Conversation}.
\newblock \bibinfo{howpublished}{\url{https://github.com/microsoft/autogen}}.
\newblock


\bibitem[Ye et~al\mbox{.}(2025)]%
        {maslab_ye_2025}
\bibfield{author}{\bibinfo{person}{Rui Ye}, \bibinfo{person}{Keduan Huang},
  \bibinfo{person}{Qimin Wu}, \bibinfo{person}{Yuzhu Cai},
  \bibinfo{person}{Tian Jin}, \bibinfo{person}{Xianghe Pang},
  \bibinfo{person}{Xiangrui Liu}, \bibinfo{person}{Jiaqi Su},
  \bibinfo{person}{Chen Qian}, \bibinfo{person}{Bohan Tang},
  \bibinfo{person}{Kaiqu Liang}, \bibinfo{person}{Jiaao Chen},
  \bibinfo{person}{Yue Hu}, \bibinfo{person}{Zhenfei Yin},
  \bibinfo{person}{Rongye Shi}, \bibinfo{person}{Bo An}, \bibinfo{person}{Yang
  Gao}, \bibinfo{person}{Wenjun Wu}, \bibinfo{person}{Lei Bai}, {and}
  \bibinfo{person}{Siheng Chen}.} \bibinfo{year}{2025}\natexlab{}.
\newblock \showarticletitle{MASLab: A Unified and Comprehensive Codebase for
  LLM-based Multi-Agent Systems}.
\newblock \bibinfo{journal}{\emph{arXiv preprint arXiv:2505.16988}}
  (\bibinfo{year}{2025}).
\newblock
\urldef\tempurl%
\url{https://arxiv.org/abs/2505.16988}
\showURL{%
\tempurl}


\bibitem[Yehudai et~al\mbox{.}(2025)]%
        {agent_eval_survey}
\bibfield{author}{\bibinfo{person}{Asaf Yehudai}, \bibinfo{person}{Lilach
  Eden}, \bibinfo{person}{Alan Li}, \bibinfo{person}{Guy Uziel},
  \bibinfo{person}{Yilun Zhao}, \bibinfo{person}{Roy Bar-Haim},
  \bibinfo{person}{Arman Cohan}, {and} \bibinfo{person}{Michal
  Shmueli-Scheuer}.} \bibinfo{year}{2025}\natexlab{}.
\newblock \showarticletitle{Survey on Evaluation of {LLM}-based Agents}.
\newblock \bibinfo{journal}{\emph{arXiv preprint arXiv:2503.16416}}
  (\bibinfo{year}{2025}).
\newblock
\urldef\tempurl%
\url{https://arxiv.org/abs/2503.16416}
\showURL{%
\tempurl}


\bibitem[Zhu et~al\mbox{.}(2025)]%
        {multiagentbench_2025}
\bibfield{author}{\bibinfo{person}{Kunlun Zhu}, \bibinfo{person}{Hongyi Du},
  \bibinfo{person}{Zhaochen Hong}, {et~al\mbox{.}}}
  \bibinfo{year}{2025}\natexlab{}.
\newblock \showarticletitle{MultiAgentBench: Evaluating the Collaboration and
  Competition of LLM Agents}. In \bibinfo{booktitle}{\emph{ACL 2025}}.
\newblock
\showeprint{2503.01935}
\urldef\tempurl%
\url{https://arxiv.org/abs/2503.01935}
\showURL{%
\tempurl}


\end{thebibliography}

\end{document}